\documentclass[twocolumn,apj,floats,aps,amsmath,amssymb,nofootinbib,numberedappendix,appendixfloats,twocolappendix]{emulateapj}
\usepackage{graphicx}
\usepackage{bm}
\usepackage{hyperref}
\hypersetup{pdfauthor=Hans Baehr}
\hypersetup{backref=true, pagebackref=true, hyperindex=true, breaklinks=true, colorlinks=true, urlcolor=blue}
\hypersetup{citecolor=blue, pagecolor=red, bookmarks=true, bookmarksopen=true}
\usepackage[english]{babel}
\usepackage{tabu}
\usepackage{amsmath}
\usepackage[caption=false, position=t, singlelinecheck=off]{subfig}

\makeatletter
\newcommand*\bigcdot{\mathpalette\bigcdot@{.5}}
\newcommand*\bigcdot@[2]{\mathbin{\vcenter{\hbox{\scalebox{#2}{$\m@th#1\bullet$}}}}}
\makeatother

\begin{document}

\citestyle{egu}
\bibliographystyle{yahapj}

\title{The Concentration and Growth of Solids in Fragmenting Circumstellar Disks}
\author{Hans Baehr$^{1,*}$}
\author{Hubert Klahr$^1$}

\affil{$^1$Max Planck Institute for Astronomy, K{\"o}nigstuhl 17, D-69117 Heidelberg, Germany}

\email{baehr@mpia.de}
\altaffiltext{*}{Member of the International Max Planck Research School for Astronomy and Cosmic Physics at the University of Heidelberg (IMPRS-HD)}

\begin{abstract}
Due to the gas rich environments of early circumstellar disks, the gravitational collapse of cool, dense regions of the disk form fragments largely composed of gas. During formation, disk fragments may attain increased metallicities as they interact with the surrounding disk material, whether through particle migration to pressure maxima or through mutual gravitational interaction. In this paper, we investigate the ability of fragments to collect and retain a significant solid component through gas-particle interactions in high-resolution 3D self-gravitating shearing box simulations. The formation of axissymmetric perturbations associated with gravitational instabilities allows particles of intermediate sizes to concentrate through aerodynamic drag forces. By the onset of fragmentation, the mass of local particle concentrations within the fragment are comparable to that of the gas component and the sebsequent gravitational collapse results in the formation of a solid core. We find that these cores can be up to several tens of Earth masses, depending on grain size, before the fragment center reaches temperatures which would sublimate solids. The solid fraction and total mass of the fragment also depend on the metallicity of the young parent protoplanetary disk, with higher initial metallicities resulting in larger fragments and larger solid cores. Additionally, the extended atmospheres of these soon-to-be gas giants or brown dwarfs are occasionally enriched above the initial metallicity, provided no solid core forms in the center and are otherwise lacking in heavier elements when a core does form.
\end{abstract}


\maketitle

\section{Introduction}
\label{sec:intro}

Gas giant planets are predominantly gaseous objects, but likely with solid central cores and overall metallicities closer to stars than terrestrial planets \citep{Helled2010,Kreidberg2014}. Recent observations of the gravitational moments of Jupiter by the Juno mission indicate that it has a substantial solid or partially dissolved core of $>10 M_{\oplus}$ \citep{Bolton2017}. Such a massive core is taken to be a strong indicator of formation by core accretion, as the accretion of a significant gas envelope requires a planetary core of around $10 M_{\oplus}$ \citep{Pollack1996}. Formation through gravitational instabilities (GI), on the other hand, involves the direct collapse of dense, gaseous regions into fragments. These massive, cool disks are most likely when a disk is young and devoid of considerable particle accumulations, neither a core or metallicity over the stellar value is expected through gravitational instabilities; however recent simulations suggest that fragments may have atmospheres which are distinct from the disk and envelope from which they form \citep{Boley2010}.

Observations of giant exoplanets and their atmospheres have led to the discovery of a number of trends in atmospheric chemical compositions and the development of theories which link these compositions to planet formation. Whether or not the solid material is sequestered in the core can affect the chemical balance of key molecules \citep{Ilee2017}. Atmospheric carbon to oxygen ratios (C/O) might be an indicator of formation location or formation mechanism \citep{Oberg2011,Brewer2017} and can be inferred through spectroscopic measurements of transiting planets \citep{Charbonneau2002} and directly imaged planets \citep{Lee2013}. While transiting planets are typically located at orbits too close to be formed \textit{in situ} by GI, they still provide a useful constraints on expected metallicity values of exoplanet atmospheres \citep{Espinoza2017,Thorngren2016}. In the case of the HR8799 system, the elemental abundances abundances are most consistent with superstellar C/O ratios in favor of core accretion, but the discrepancy between inner and outer planets suggests different accretion histories and formation pathways are possible \citep{Lavie2017}.

\begin{figure*}[t]
\centering
\includegraphics[width=0.5\textwidth]{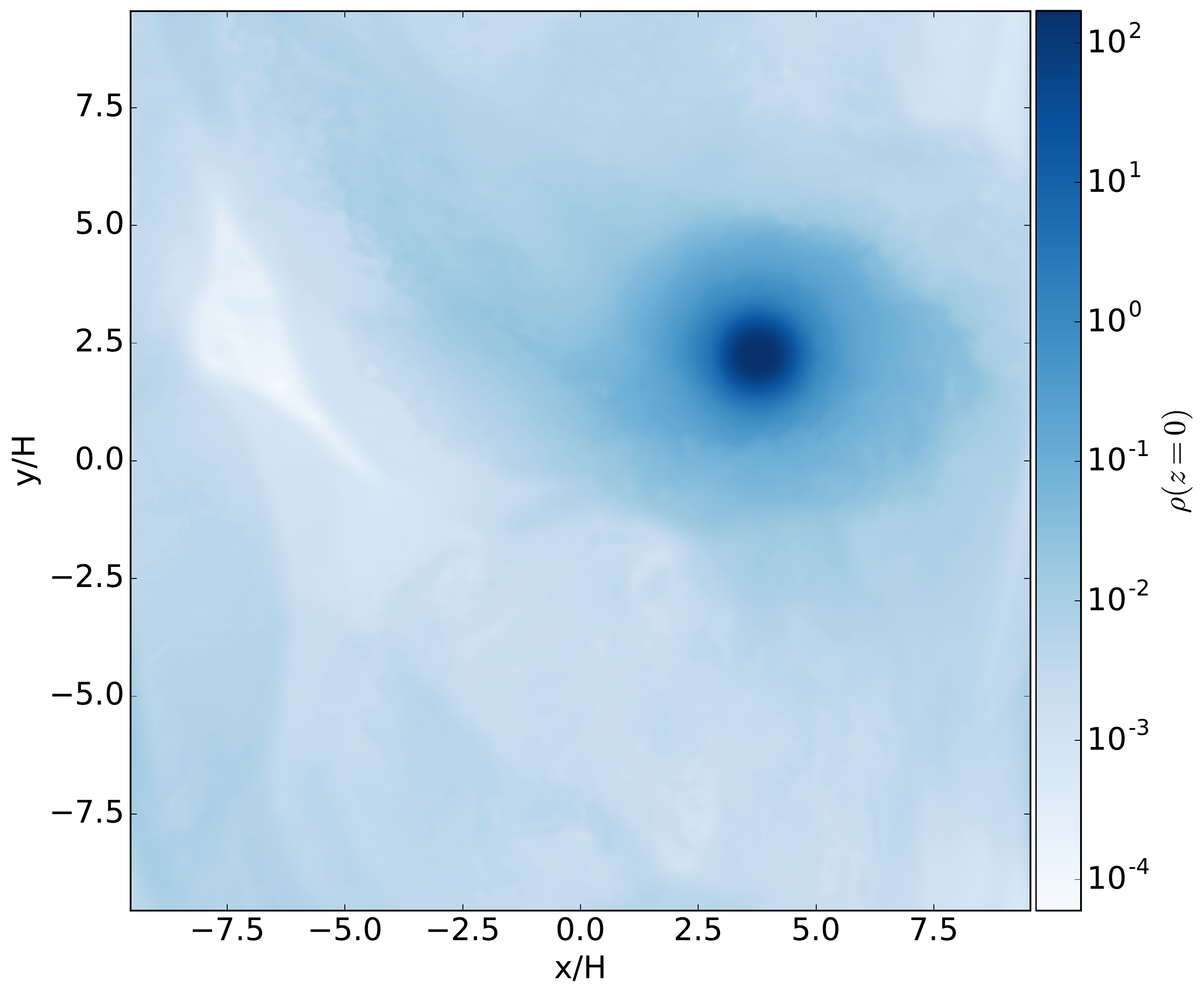}%
\includegraphics[width=0.5\textwidth]{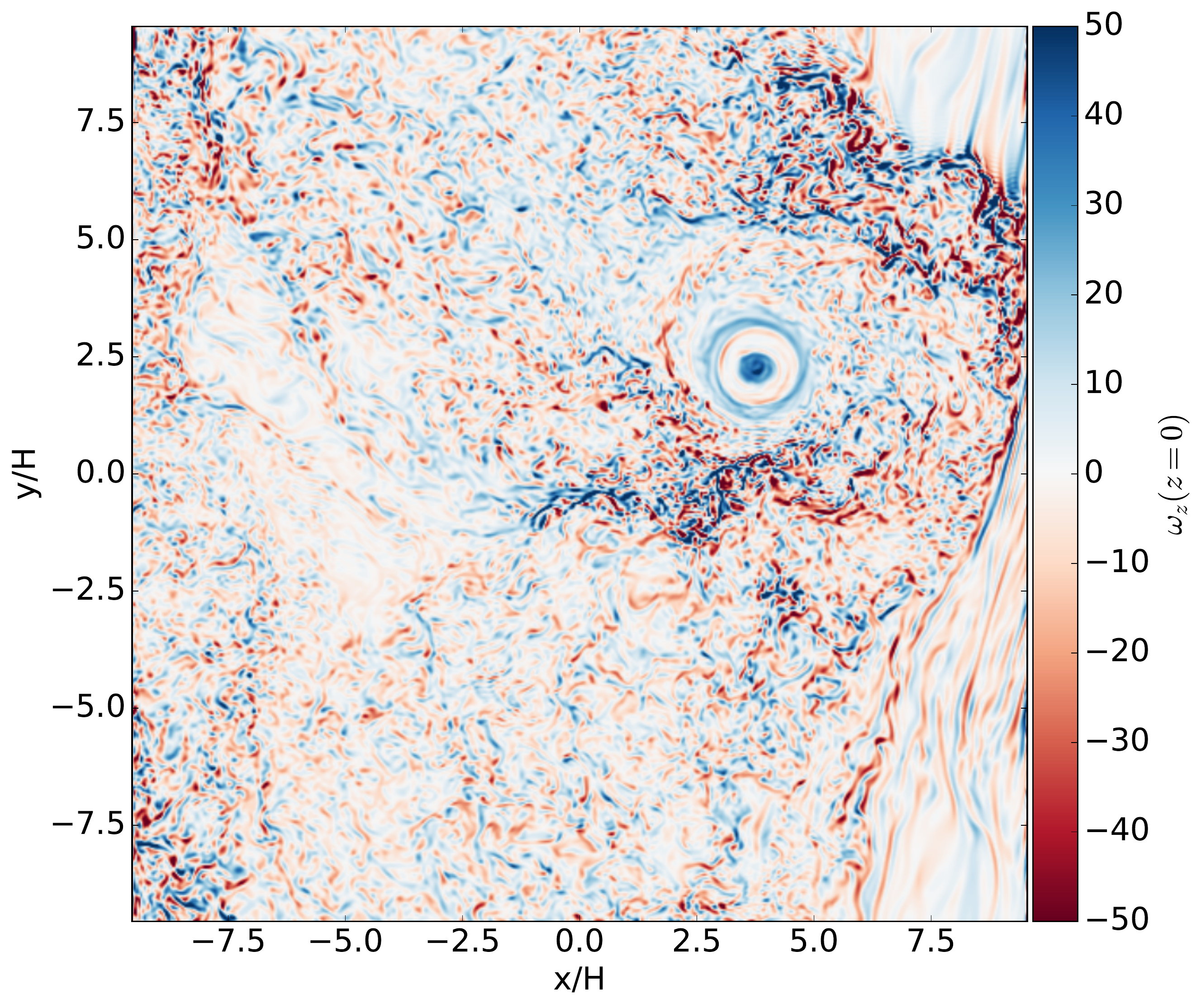}
\caption{A slice through the midplane of the gas density (left) and vertical component of the vorticity (right) of a fragmenting 3D simulation without particles at $512^{2} \times 64$ resolution. The vorticity gives a sense of the turbulent flows which are not readily apparent in the gas density, including differential rotation of the fragment. These simulations are indicative of the gas fragmentation behavior without uniformly-sized feedback particles.}
\label{fig:3d}
\end{figure*}

Gravitationally unstable disks may still be capable of producing large gaseous objects with significant solid cores by considering that even in low dust-to-gas environments, dust can collect in fragments \citep{Helled2006}. While spiral arms produced by GI are well documented locations of particle concentration \citep{Gibbons2012,Boss2015}, it is still unknown just how enriched a fragment formed therein may become enriched with solids during formation and immediately after, when temperatures are not high enough to evaporate solids. \citet{Boley2010} investigated particle concentration in self-gravitating disks with global simulations that included two particle sizes, and found considerable solid concentrations (i.e. cores and embryos), interior and exterior to fragments. With high-resolution shearing box simulations of fragmentation, we aim to explore this process further.

In the linear analysis of the gravitational collapse of a gaseous disk, stability is defined according to the Toomre parameter \citep{Toomre1964,Goldreich1965}
\begin{equation} \label{eq:toomreparameter}
Q = \frac{c_{\textnormal{s}}\Omega}{ \pi G\Sigma},
\end{equation}
where values above $Q=1$ represent marginal stability to fragmentation, but non-axisymmetric structures may form up to $Q \sim 1.5$ \citep{Durisen2007}. Here, $c_{\textnormal{s}}$ is the isothermal sound speed and $\Omega = (GM/R^3)^{1/2}$ is the orbital frequency corresponding to the stabilizing effects of thermal pressure and tidal shear respectively, which balance the gravitational collapse of a significantly overdense mass surface density $\Sigma$, with constants $\pi$ and gravitational constant $G$. The formation of fragments in nonlinear gravitoturbulent evolution requires the fulfillment of the Gammie cooling criterion, whereby the disk cools on sufficiently short timescales such that thermal pressure support is removed on approximately the orbital timescale $t_{\mathrm{c}} = \beta\Omega^{-1}$, with a critical value around $\beta_{\mathrm{crit}} = 3$ \citep{Gammie2001,Deng2017,Baehr2017}.

A number of studies have looked into the behavior of solid particles in marginally gravitationally unstable disks, but have mostly focused on particle evolution in gravitoturbulent disks. These range from looking into the collection of particles in voritices \citep{Gibbons2015} and fragments \citep{Boley2010}, shock fronts \citep{Gibbons2012}, and other relations with the GI turbulence \citep{Shi2016}. These find that the particles of typically moderate sizes will collect and form high solid concentrations through aerodynamic drag and gas-dust coupling. \citet{Shi2013} consider a concentration of particles well above the typical local dust-to-gas ratios for gas disks so that the particles will naturally fragment even if the gas disk is not gravitationally unstable (i.e. $Q_{\rm{gas}} \geq 10$). We are interested in the regime where the gas disk is gravitationally unstable ($Q_{\rm gas} \leq 1$), but the dust is not, therefore the gas will fragment but initial dust overdensities are not formed through the direct gravitational collapse of the dust, but by the hydrodynamic interactions between the gas and the dust.

For the study of fragment formation including particles, \citet{Boley2010} used global radiation hydrodynamic simulations, investigating the large-scale structure of gravitationally unstable disks with two particle sizes. In these simulations, particles are first concentrated into the spiral arms, reaching surface densities which start contributing to the self-gravity in these regions, warranting the inclusion of particle self-gravity. At this point, the particles and gas then collapse together, and the particles rapidly concentrate at the center and potentially constitute a new mode of planet formation \citep{Nayakshin2014}. \citet{Boley2010} ultimately find that there are significant concentrations of small grains at the center of fragments, however, their resolution is insufficient to capture the effects of small scale turbulence caused by parametric instabilities near the midplane \citep{Riols2017}. 

\begin{table*}[t]
\caption{\textnormal{Performed simulations and their various resolutions, particle sizes St, initial metallicities $Z_{0}$, Fragment metallicity $Z/Z_{0}$, fragment and core masses}}
\centering  
\begin{tabular}{l*{6}{c}r}
Simulation  & Grid Cells & Particle number & Stokes \# & $Z_{0}$ & $Z/Z_{0}$ & Fragment mass $M_{Jup}$ & Core mass $M_{\oplus}$ \\
\hline
G512t2             & $512^2 \times 64$  &  $-$   & $-$     & $-$      &   $-$ & 14.8 & $-$ \\

P512t2pss          & $512^2 \times 64$  &  $10^6$& $0.01$  & $10^{-2}$&  3.4  & 5.24 & 57.9\\
P512t2ps           & $512^2 \times 64$  &  $10^6$& $0.1$   & $10^{-2}$&  14.6 & 1.32 & 62.3\\
P512t2p            & $512^2 \times 64$  &  $10^6$& $1$     & $10^{-2}$&  1.8  & 16.2 & 89.1\\
P512t2pl           & $512^2 \times 64$  &  $10^6$& $10$    & $10^{-2}$&  1.7  & 2.3  & $-$ \\

P1024t2pss         & $1024^2 \times 128$  &  $10^6$& $0.01$ & $10^{-2}$&  1.7 & $-$  & $-$ \\
P1024t2ps          & $1024^2 \times 128$  &  $10^6$& $0.1$  & $10^{-2}$&  5.1 & 4.8  & 69.0\\
P1024t2p           & $1024^2 \times 128$  &  $10^6$& $1$    & $10^{-2}$&  3.8 & 5.7  & 70.7\\
P1024t2pl          & $1024^2 \times 128$  &  $10^6$& $10$   & $10^{-2}$&  1.5 & 7.19 & $-$ \\

P1024t2psl         & $1024^2 \times 128$  &  $10^6$& $0.1$  & $8\times 10^{-3}$   &  5.0 & 3.83 & 49.3 \\
P1024t2psm         & $1024^2 \times 128$  &  $10^6$& $0.1$  & $1.33\times 10^{-2}$&  4.8 & 8.01 & 147.6\\
\end{tabular}
\label{tab:sims}
\end{table*}

Since young gravitationally unstable disks are largely dominated by their gas content, the distribution and evolution of solids can often be overlooked or ignored for convenience. A handful of studies have looked into the interaction between the gas and dust in gravitoturbulent disks \citep{Gibbons2012,Gibbons2014a,Gibbons2015,Shi2013,Shi2016}, and have found varying degrees of solid concentration within disk features such as spirals arms, shocks and vortices. However, none of these considered the effect of a fragmenting disk on particle distribution and concentration. Thus this study will focus on the evolution of solids in a disk in which the gas is gravitationally unstable, but the dust, being at a realistic metallicity of the interstellar medium, is not initially susceptible to direct gravitational collapse. This may help determine the initial metallicities of gas giant planets formed though disk instability as the size and mass of any significant solid clumps. In this paper we will use
\begin{equation} \label{eq:metallicity}
Z \equiv M_{\rm solid}/M_{\rm gas},
\end{equation}
to denote the mass ratio of solid to gas, within the entire simulation domain, unless otherwise stated. For the local dust-to-gas density ratio within a single grid cell we use 
\begin{equation} \label{eq:dusttogasratio}
\epsilon \equiv \rho_{\rm d}/\rho_{\rm g}.
\end{equation}
The vorticity of the gas is a measure of the local rotation of a fluid and is a simple diagnostic of turbulent motion, defined as the curl of the velocity field $\bm{u}$: $\omega = \nabla\times\bm{u}$. In the case of our 2D midplane figures, the value of the vertical component is plotted, which also serves to show the rotation in and around a fragment
\begin{equation} \label{eq:vorticity}
\omega_{z} = \left( \frac{\partial u_{y}}{\partial x} - \frac{\partial u_{x}}{\partial y} \right) \bm{\hat{z}}.
\end{equation}
Post-formation enrichment was studied by \citet{Helled2008} and \citet{Helled2010}, assuming that fragments are formed initially as pure gas overdensities which subsequently accrete nearby material. While they find enrichment of the atmosphere is possible through the accretion of planetesimals, core formation is excluded due to the high temperatures at the center which evaporate all solids which do accrete and convection, which prevents dust settling. However, sublimation temperatures at the fragment core are only reached after tens of thousands of years, a large enough interval such that particles accumulated during formation may settle and form a core by the time sublimation occurs \citep{Nayakshin2010a}. Convection is less of a hindrance in the distant protoplanetary disk ($\sim 100$ au) where densities and optical depths are lower \citep{Nayakshin2010a}. This is the primary focus of the present investigation: attempting to quantify how much material is present immediately after fragment formation which could potentially survive sublimation at the center. However, additional attention will also be paid to the solid content in the atmospheres and its consequences regarding directly observed planets and their formation.

We continue by introducing the important hydrodynamic equations and processes necessary to describe a self-gravitating particle-gas mixture in Section \ref{sec:theory} followed by a brief description of the numerical implementation of particles and initial conditions in Section \ref{sec:models}. In Section \ref{sec:results} we present the results of our simulations of 3D gravitationally unstable disks with particles of various sizes, including estimates of core masses and fragment metallicities. We conclude by discussing the implications on planet formation theory and observations in Sections \ref{sec:discussion} and \ref{sec:conclusion}.

\section{Theory}
\label{sec:theory}

For this study we conduct 3D hydrodynamic shearing box simulations of a self-gravitating disk with Lagrangian 'swarm' particles embedded in the Eulerian mesh using the \textsc{Pencil} \footnote{http://pencil-code.nordita.org/} code. Both gas and particles are treated as self-gravitating and particle back-reaction is calculated on the gas by assuming particles are mapped to the grid with triangular-shaped clouds. Local simulations allow the Toomre wavelength $\sim 2\pi H$ to be well-resolved in the radial and azimuthal coordinates ($x$ and $y$ in the linearized coordinates, respectively) and keep the boundary conditions periodic. This Toomre wavelength is not resolved in the vertical direction $z$, but this has little consequence on fragmentation, which is dominated by radial and azimuthal collapse. Shearing box simulations use hydrodynamic equations which are linearized and transformed into co-rotating Cartesian coordinates, where $q = \mathrm{ln}\Omega / \mathrm{ln}R = 3/2$ is the shear parameter:

\begin{align}
\frac{\partial {\rho_{\mathrm{g}}}}{\partial t} &- q\Omega x\frac{\partial {\rho_{\mathrm{g}}}}{\partial y} + \nabla\cdot(\rho_{\mathrm{g}}\bm{u}) = f_{D}(\rho_{\mathrm{g}}) \label{eq:finalmassconserve} \\
\frac{\partial \bm{u}}{\partial t} &- q\Omega x\frac{\partial \bm{u}}{\partial y} + \bm{u}\cdot\nabla\bm{u} = -\frac{\nabla P}{\rho_{\mathrm{g}}} + q\Omega v_{x}\bm{\hat{y}} \nonumber \\ 
& - 2\Omega\times\bm{u} - \nabla\Phi + \frac{\epsilon}{\tau_{s}} ( \bm{u} - \bm{w} ) + f_{\nu}(\bm{u}) \label{eq:finalmomconserve} \\
\frac{\partial s}{\partial t} &-q\Omega x\frac{\partial s}{\partial y} + (\bm{u} \cdot \nabla)s = \nonumber \\
& \frac{1}{\rho_{\mathrm{g}} T} \left( 2\rho_{\mathrm{g}}\nu\mathbf{S}^{2} - \Lambda + f_{\chi}(s) \right). \label{eq:finalenergyconserve}
\end{align}
In equations \eqref{eq:finalmassconserve} - \eqref{eq:finalenergyconserve}, $\mathbf{u} = (v_{\textnormal{x}},v_{\textnormal{y}}+q\Omega x,v_{\textnormal{z}})^{T}$ is the gas flow plus shear velocity in the local box, $\bm{w}$ is the particle velocity which imparts a back reaction on to the gas proportional to the local dust-to-gas ratio $\epsilon$, $\rho_{\mathrm{g}}$ is the gas density, and the thermodynamic variable is the gas entropy $s$. Viscous heat is generated by $\mathcal{H} = 2\rho_{\mathrm{g}}\nu\mathbf{S}^{2}$, with rate-of-strain tensor $\mathbf{S}$. Hyperdissipation is applied with the terms $f_{D}(\rho_{\mathrm{g}})$, $f_{\nu}(\bm{u})$, $f_{\chi}(s)$ which for each has the form
\begin{equation} \label{eq:hyperdiff}
f(\xi) = \nu(\nabla^{6}\xi),
\end{equation}
with constant $\nu = 2.5\, H^{6}\Omega$ \citep{Yang2012}. Heat is lost via simple $\beta$-cooling prescription
\begin{equation} \label{eq:coolingfunction} 
\Lambda = \frac{\rho (c_{\textnormal{s}}^{2} - c_{\textnormal{s,irr}}^{2})}{(\gamma -1) t_{\textnormal{c}}} 
\end{equation}
with $t_{\textnormal{c}}$ given by $t_{\mathrm{c}} = \beta\Omega^{-1}$ and background irradiation term $c_{\textnormal{s,irr}}^{2}$, which ensures fragmentation is a result of growing mass perturbations rather than a transient region of effectively no thermal pressure support. This cooling prescription has no dependence on the optical depth and thus all regions cool with the same efficiency. In reality, the opacity will be dominated by the dust grains and an increase of the particle density will increase the local cooling timescale.

Self-gravity is solved in Fourier space by transforming the surface density to find the potential at wavenumber $k$ and transforming the solution back into real space. The solution to the Poisson equation in Fourier space at wavenumber $k$ is
\begin{equation} \label{eq:gravpotential}
\Phi(\bm{k}, t) = -\frac{2\pi G\rho(\bm{k}, t)}{\bm{k}^2},
\end{equation}
where $\Phi = \Phi_{\mathrm{g}} + \Phi_{\mathrm{d}}$ and $\rho = \rho_{\mathrm{g}} + \rho_{\mathrm{d}}$ are the potential and density of the gas plus dust particles combined.

Finally, we use an ideal equation of state, with internal energy $\varepsilon$, and specific heat ratio $\gamma$
\begin{equation} \label{eq:eos}
P = (\gamma - 1)\rho\varepsilon.
\end{equation}

\begin{figure}[t]
\centering
\includegraphics[width=0.5\textwidth]{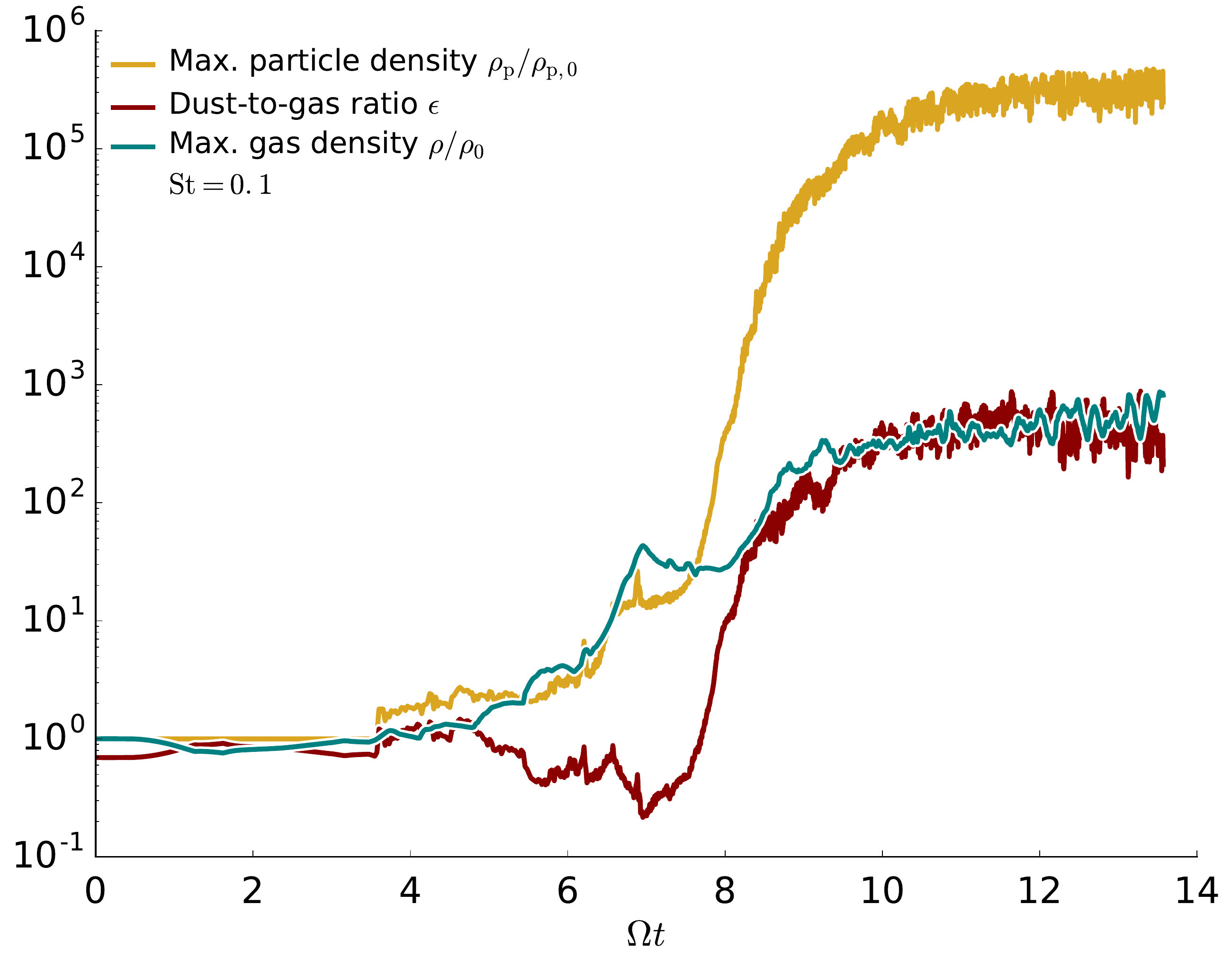}
\includegraphics[width=0.5\textwidth]{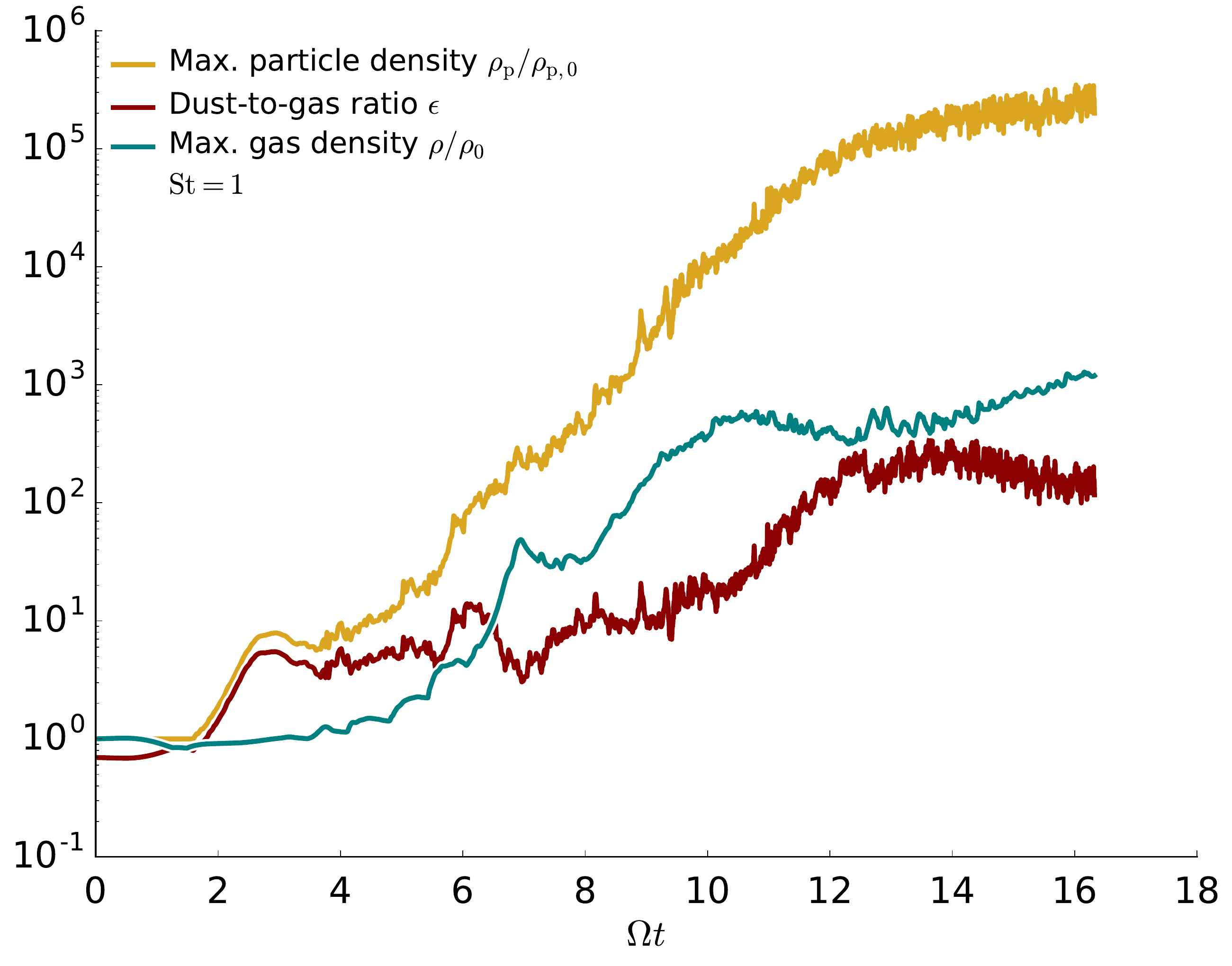}
\includegraphics[width=0.5\textwidth]{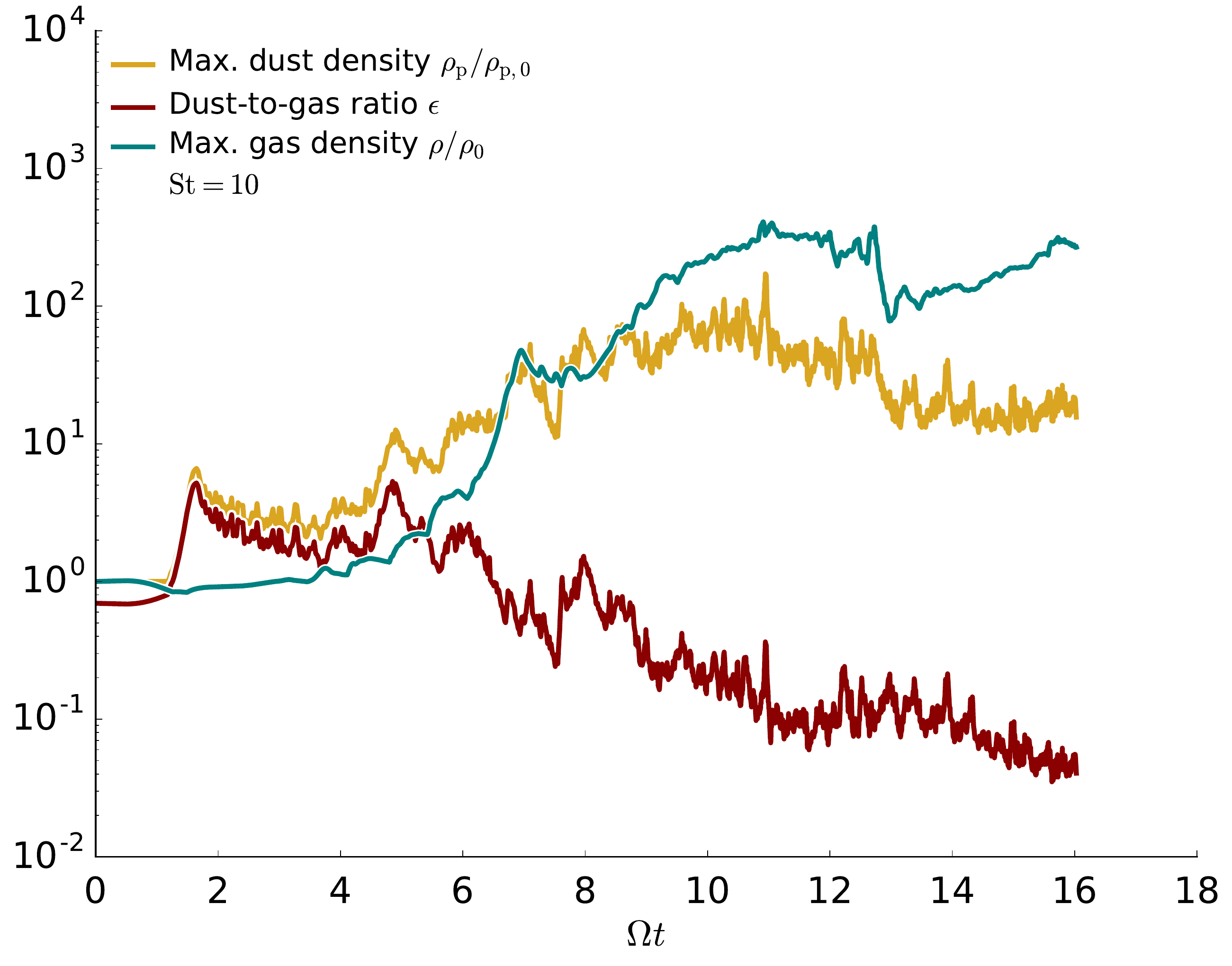}
\caption{Time evolution of the maximum particle density (yellow lines), maximum gas density (teal lines) and dust-to-gas ratio $\epsilon$ (red lines) for the particle stokes numbers in our 3D simulations: $\mathrm{St}=0.1$ (top), $\mathrm{St}=1$ (middle),  and $\mathrm{St}=10$ (bottom). The value of $\epsilon$ is that of the center of the clump, which corresponds to the peak particle and gas densities in the simulation.}
\label{fig:3ddensityevolution}
\end{figure}

Particles are added such that the initial distribution maintains a physically motivated metallicity of 1:100, roughly that of the interstellar medium (ISM), although this may be an overestimate of the gas content of the disk \citep{Ansdell2016,Miotello2017}. Stars may form from clouds of a wide range of metallicities and this may have an impact on planet formation, so deviations from this value will be value will be investigated later. Particles evolve as a collection of solids, also known as a superparticle, $i$, with position $\bm{x}^{(i)}$ and velocity $\bm{w}^{(i)}$ as in \citet{Youdin2007}
\begin{align}
\frac{d \bm{w}^{(i)}}{d t} &= 2\Omega w_{y}^{(i)} \bm{\hat{x}} - \frac{1}{2} \Omega w_{x}^{(i)} \bm{\hat{y}} - \nabla\Phi \nonumber \\
&+ \frac{1}{\tau_{s}} \left( \bm{w}^{(i)} - \bm{u}(\bm{x}^{(i)}) \right) \label{eq:particlevel}\\
\frac{d \bm{x}^{(i)}}{d t} &= \bm{w}^{(i)} - \frac{3}{2}\Omega x^{(i)} \bm{\hat{y}}, \label{eq:particlepos}
\end{align}
where $\tau_{f}$ (also referred to as the Stokes number) is the particle stopping time normalized by the dynamical time $\Omega^{-1}$
\begin{equation} \label{eq:stokenum}
\mathrm{St} = \tau_{f} = \tau_{\mathrm{s}}\Omega.
\end{equation}

\begin{figure*}[t]
\centering
\includegraphics[width=0.5\textwidth]{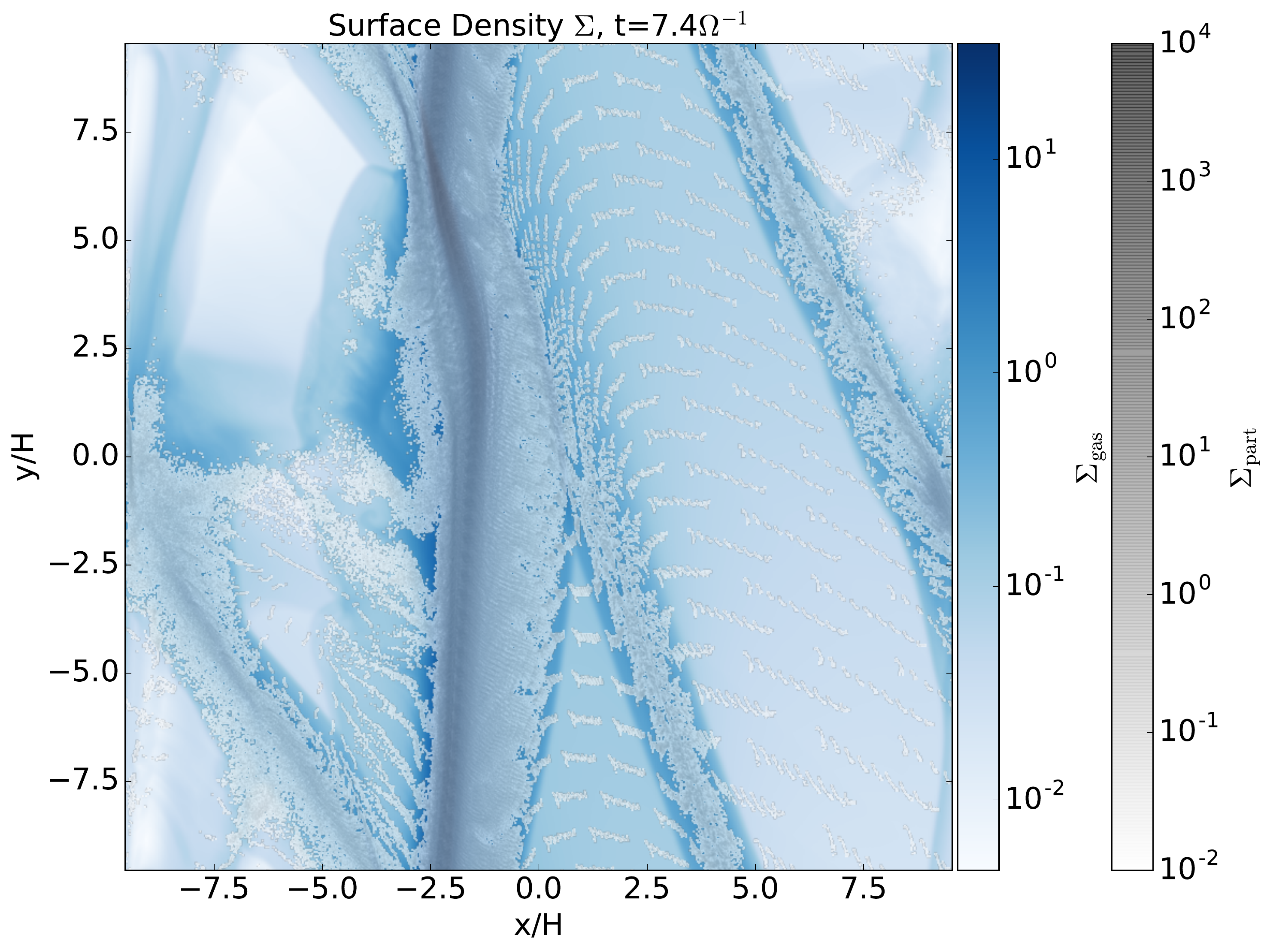}%
\includegraphics[width=0.5\textwidth]{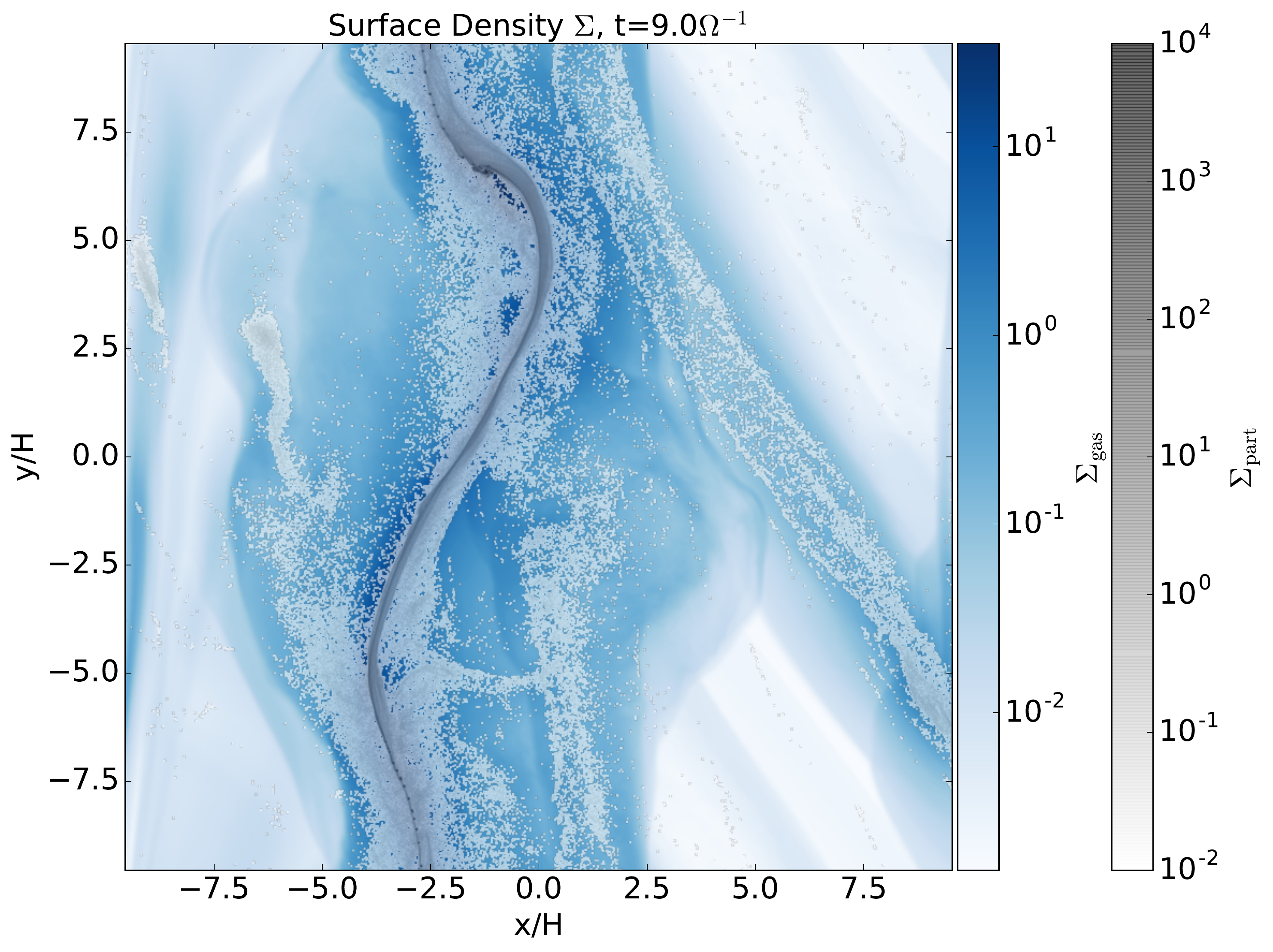}
\includegraphics[width=0.5\textwidth]{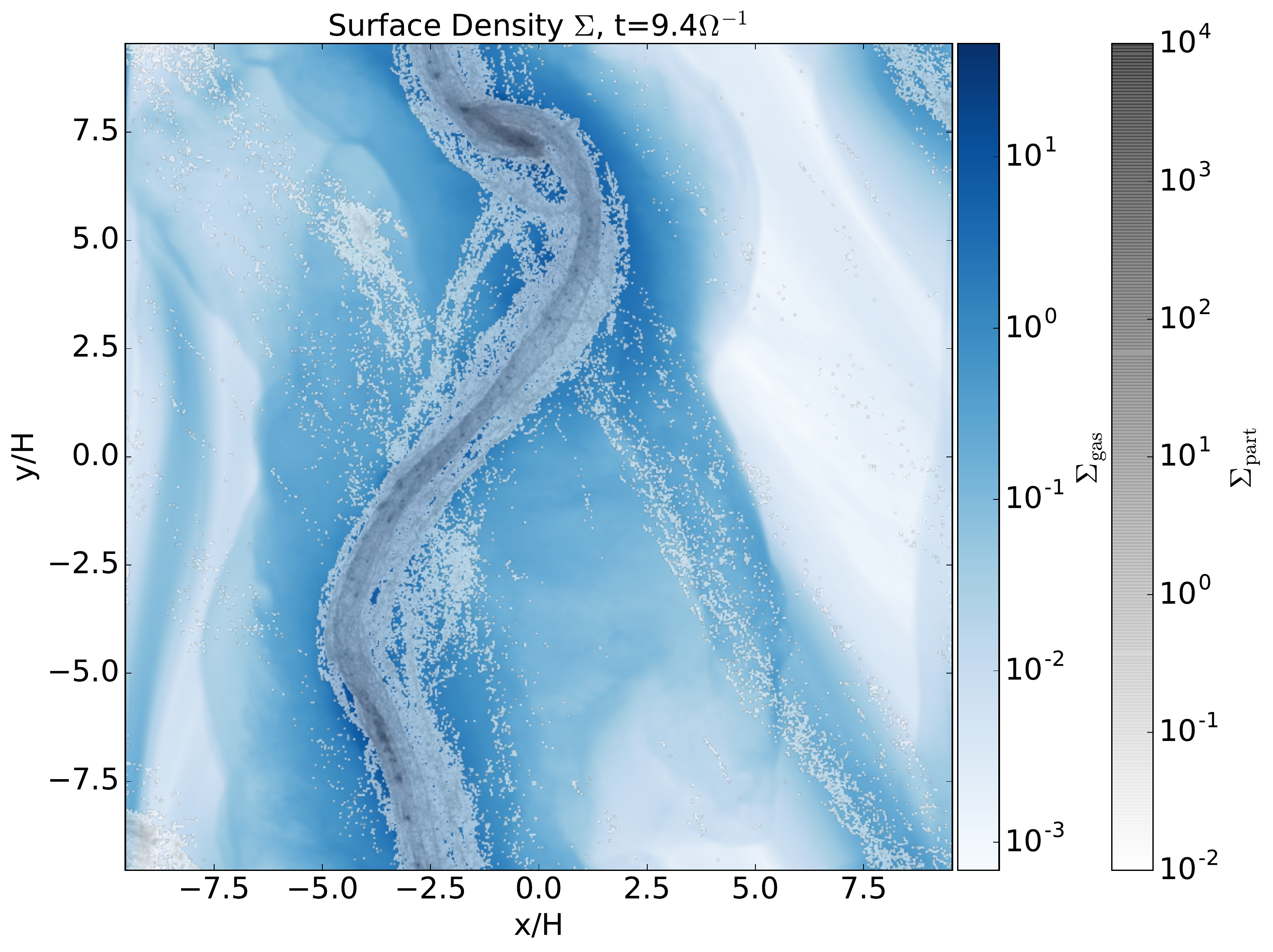}%
\includegraphics[width=0.5\textwidth]{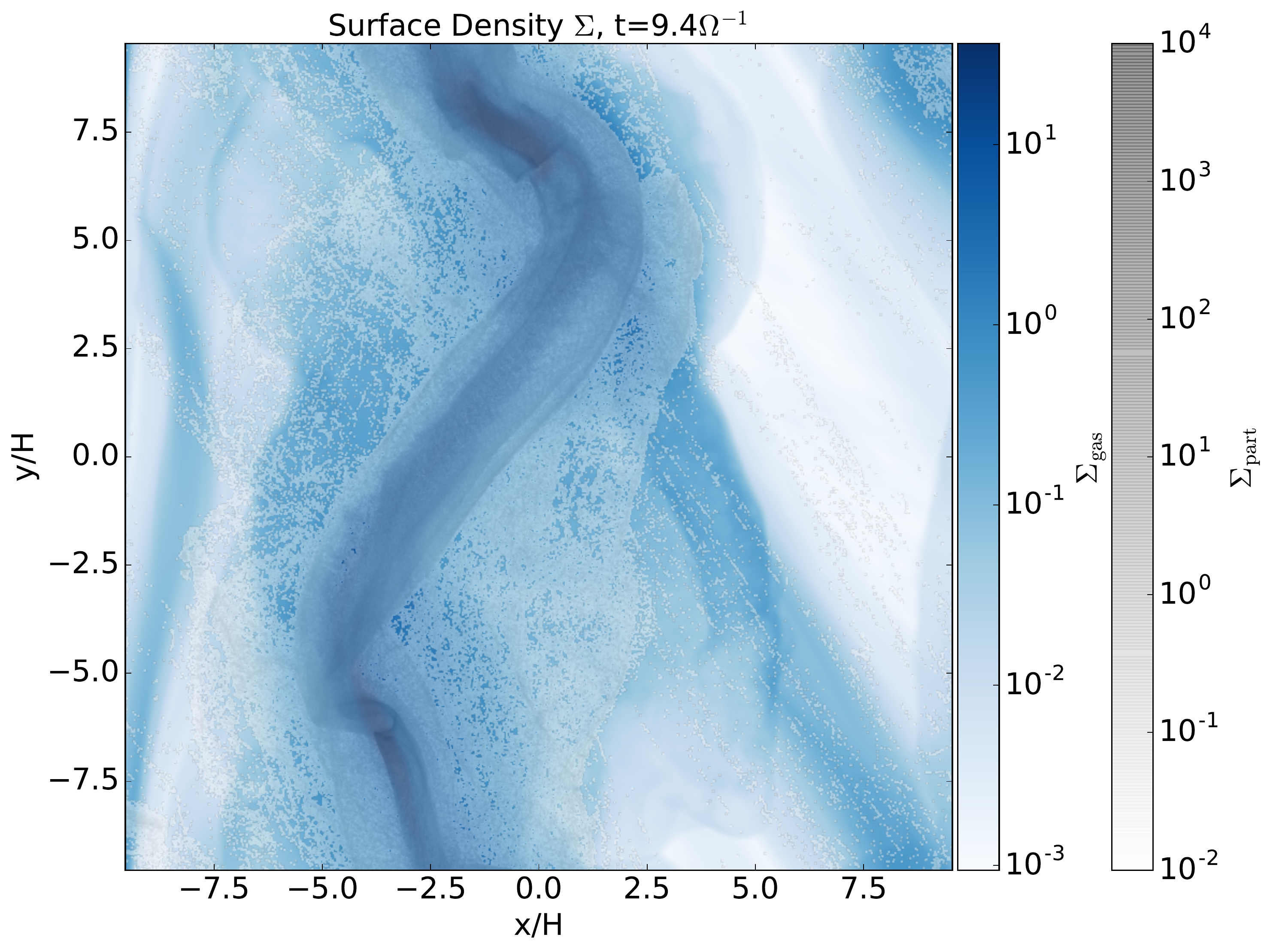}%
\caption{Map of the gas surface density (blue) with particle surface density superimposed at roughly the time the simulation transitions from forming the linear axisymmetric perturbation to the non-linear collapse of the fragment. It is at this stage that particles are being concentrated into densities which begin affecting the total self-gravitational potential of the system. Each simulation has a million particles of a single species: $\mathrm{St}=0.01$ (top left), $\mathrm{St}=0.1$ (top right), $\mathrm{St}=1.0$ (bottom left) and $\mathrm{St}=10$ (bottom right). Particle sizes $\mathrm{St}=0.1$ and $\mathrm{St}=1.0$ will concentrate enough to considerable solid cores immediately, while $\mathrm{St}=0.01$ and $\mathrm{St}=10$ will have marginal solid concentrations. Particle striation visible in the upper right panel is a result of the initial conditions explained in Section \ref{sec:results}.}
\label{fig:particlemap}
\end{figure*}

This gives a hydrodynamic sense of the particle size in the Epstein regime, where the stopping time is defined as \citep{Weidenschilling1977}
\begin{equation} \label{eq:epsteinregime}
\tau_{\mathrm{s}} = \frac{a\rho_{\bigcdot}}{c_{\mathrm{s}}\rho_{\mathrm{g}}}
\end{equation}
where $a$ is the particle diameter and $\rho_{\bigcdot}$ is the material density of a superparticle. larger particles are less coupled to small scale gas motions, retain their initial perturbations for longer and thus have higher Stokes numbers. As a corollary, smaller particles are well-coupled and will quickly take the form of the gas motions in its vicinity and thus have low Stokes numbers. Particle mass is calculated from the gas through the initial condition of the metallicity.

We include no radial pressure gradient so the radial migration of particles is not included. Because we include particles at the ISM dust-to-gas ratio $\epsilon = 10^{-2}$ the contribution of particles to the gravitational potential is initially miniscule and thus the potential is dominated by the gas distribution. All simulations include a million particles which are initially uniformly distributed throughout the domain, but rapidly settle to the disk midplane and concentrate with the gas overdensities. Because the gas is stratified and the particles are not, the initial dust-to-gas values near the vertical boundaries will be higher than the overall metallicity of $1/100$ and the midplane values will be lower although the overall metallicity is still at the desired value.

\section{Numerical Model}
\label{sec:models}

All simulations were run using the high-order finite-difference code {\scshape Pencil} \citep{Brandenburg2003} well-suited for turbulent flows. The code is stabilized through a six-order hyperdissipation scheme \citep{Yang2012} ensuring power is preserved at large scales, but numerical noise damped at small scales. The 3D gas-only simulation shown in Figure \ref{fig:3d} and included here for reference, is based on those in \citet{Baehr2017} but altered to match the dimensions of the other runs presented here; more detailed descriptions of the numerical methods and initial conditions can be found therein.

\begin{figure*}[t]
\centering
\includegraphics[width=0.5\textwidth]{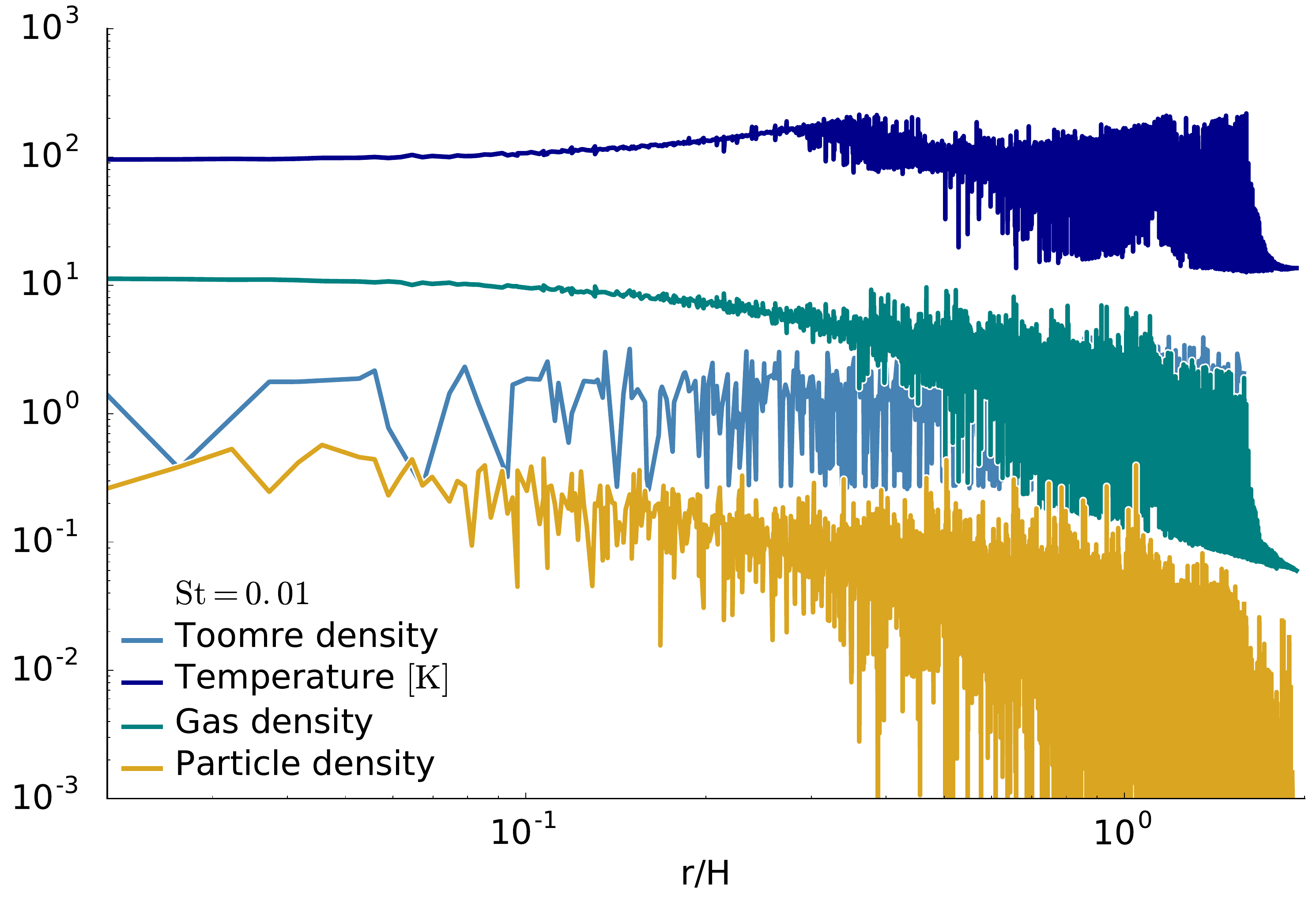}%
\includegraphics[width=0.5\textwidth]{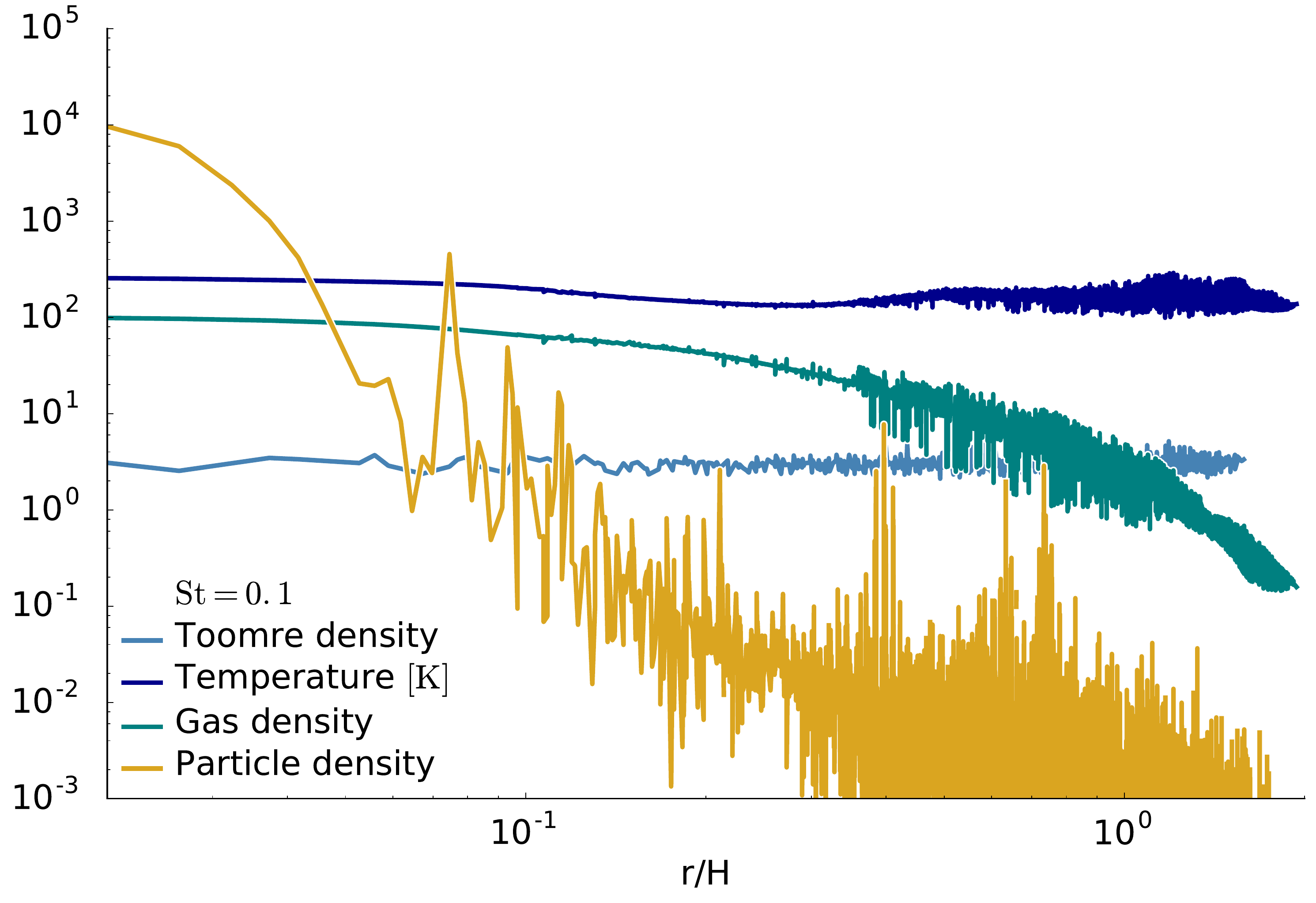}
\includegraphics[width=0.5\textwidth]{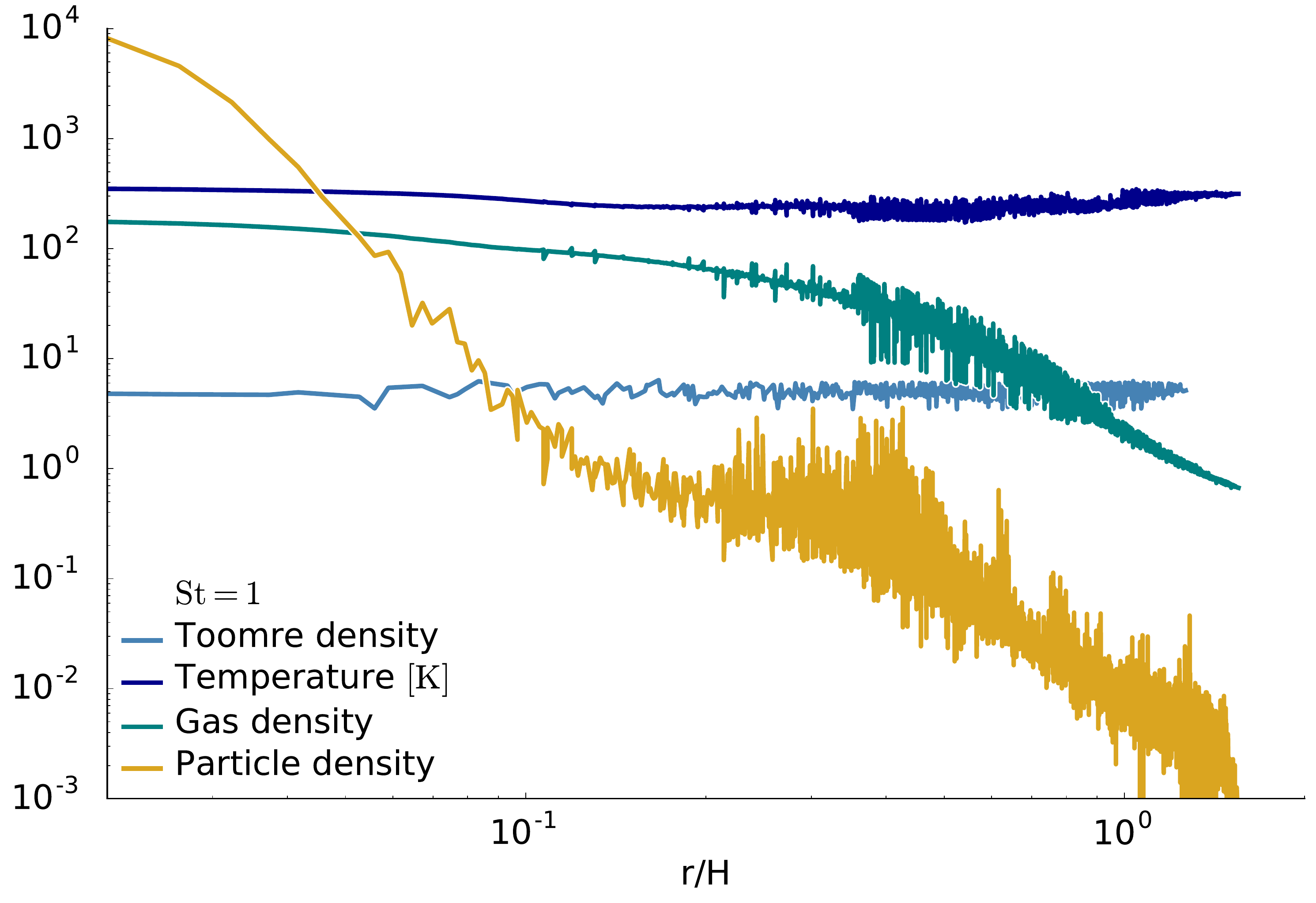}%
\includegraphics[width=0.5\textwidth]{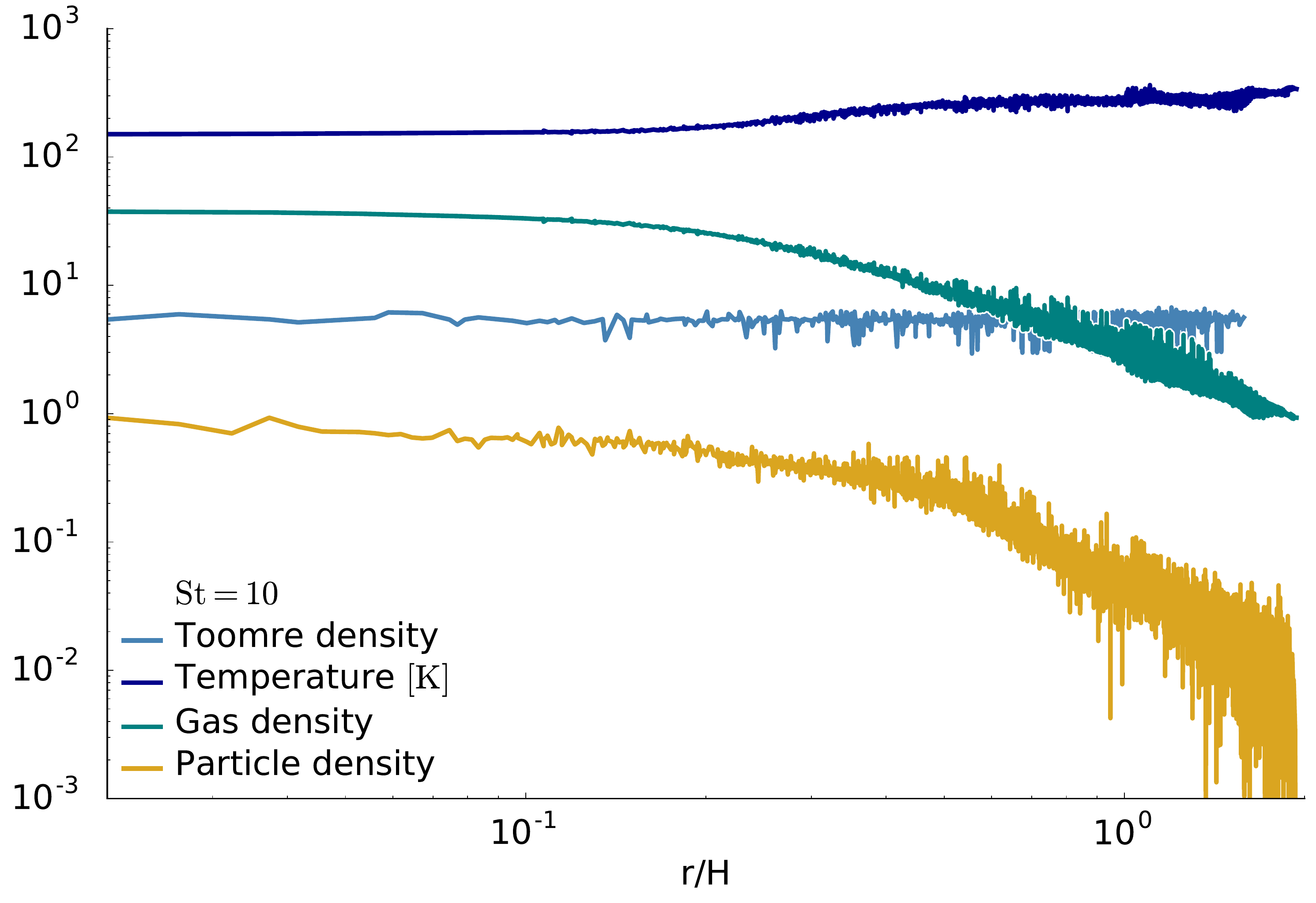}%
\caption{Radial profile of a fragment by gas density, particle density and local Toomre density (all in code units) and temperature (in $\rm K$) for simulations containing a single particle species of size $\mathrm{St}=0.01$ (top left), $\mathrm{St}=0.1$ (top right), $\mathrm{St}=1.0$ (bottom left) and $\mathrm{St}=10$ (bottom right). Values are determined by first locating the cell with the highest gas volume density and then averaging the separately the particle and gas densities for each radial distance away from that cell. The Toomre density is defined through the 3D Toomre parameter (Equation \eqref{eq:toomrevolumedensity})}
\label{fig:fragmentprofiles}
\end{figure*}

\subsection{Particles in {\scshape Pencil}}
\label{subsec:particles}

Particles are included in the {\scshape Pencil} code using Lagrangian 'swarm' particles which represent collections of similar-sized particles all moving together \citep{Johansen2007,Youdin2007,Schreiber2018}. These particles feel the effect of self-gravity and feel drag forces from the gas as well as feedback of the particles onto the gas but do not collide, merge or otherwise interact with the gas or each other. We include particles of various sizes, i.e. with a range of Stokes numbers $\mathrm{St} = [0.01, 0.1, 1, 10]$ that also quantify how quickly the particle couples to the gas, to evaluate the ability of these various sizes to be captured by fragments.

The evolution of the particles is governed by equations \eqref{eq:particlevel} and \eqref{eq:particlepos}, and are mapped to the grid using a triangle-shaped-cloud method, which distributes particle mass with a weight function $W_{\mathrm{I}} (\bm{x}^{(i)} - \bm{x})$ which puts most of the mass in the nearest cell and the rest in the 26 surrounding cells \citep{Youdin2007}. This keeps the particles from becoming too discretely distributed and for the correct calculation of the back-reaction of the gas onto the dust, which uses the same distribution function to weight the velocity contributions towards the superparticle from each of the 27 cells \citep{Yang2016}.

\subsection{Initial Conditions}
\label{subsec:initial}

\begin{figure*}[t]
\centering
\includegraphics[width=0.5\textwidth]{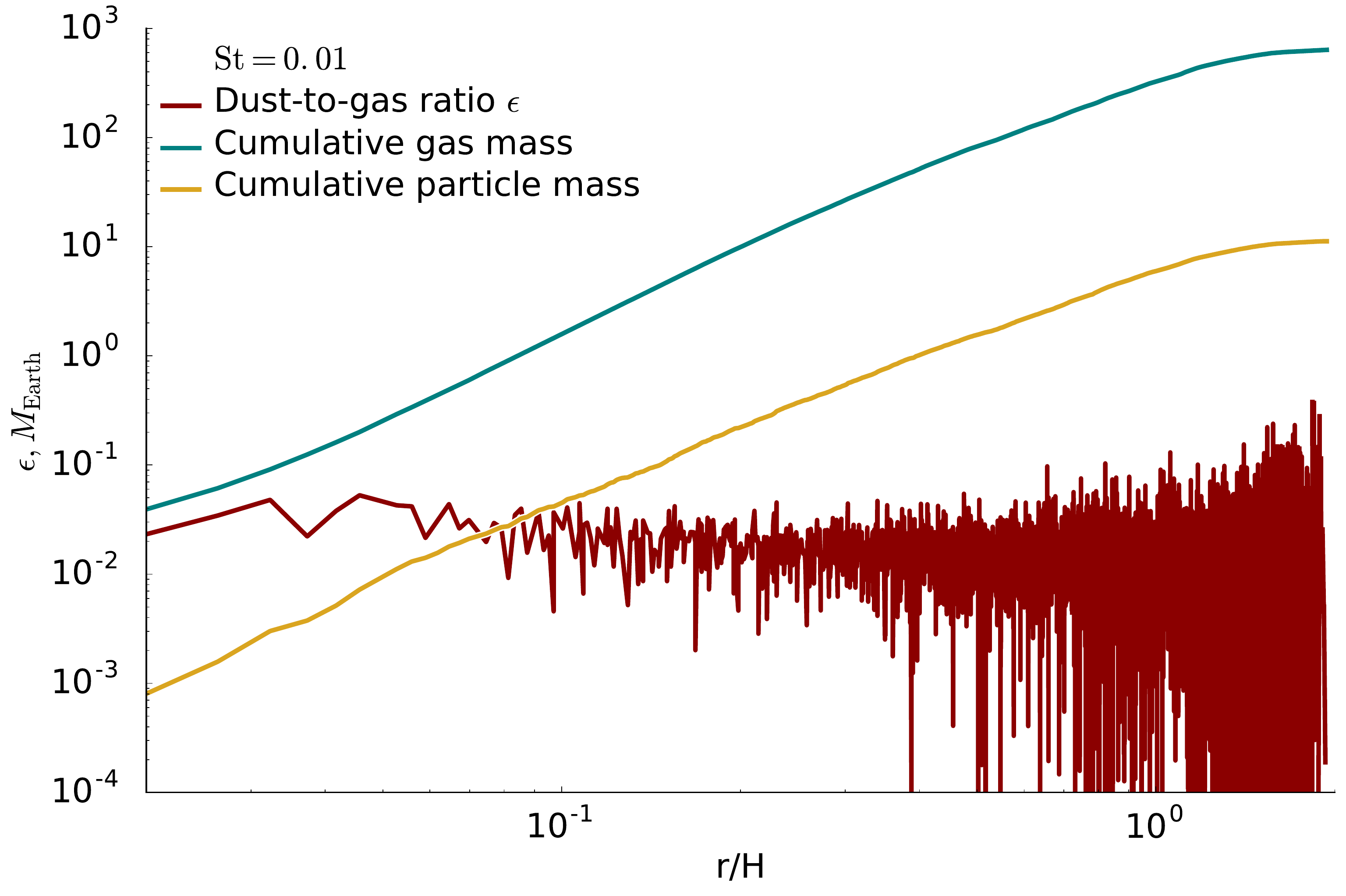}%
\includegraphics[width=0.5\textwidth]{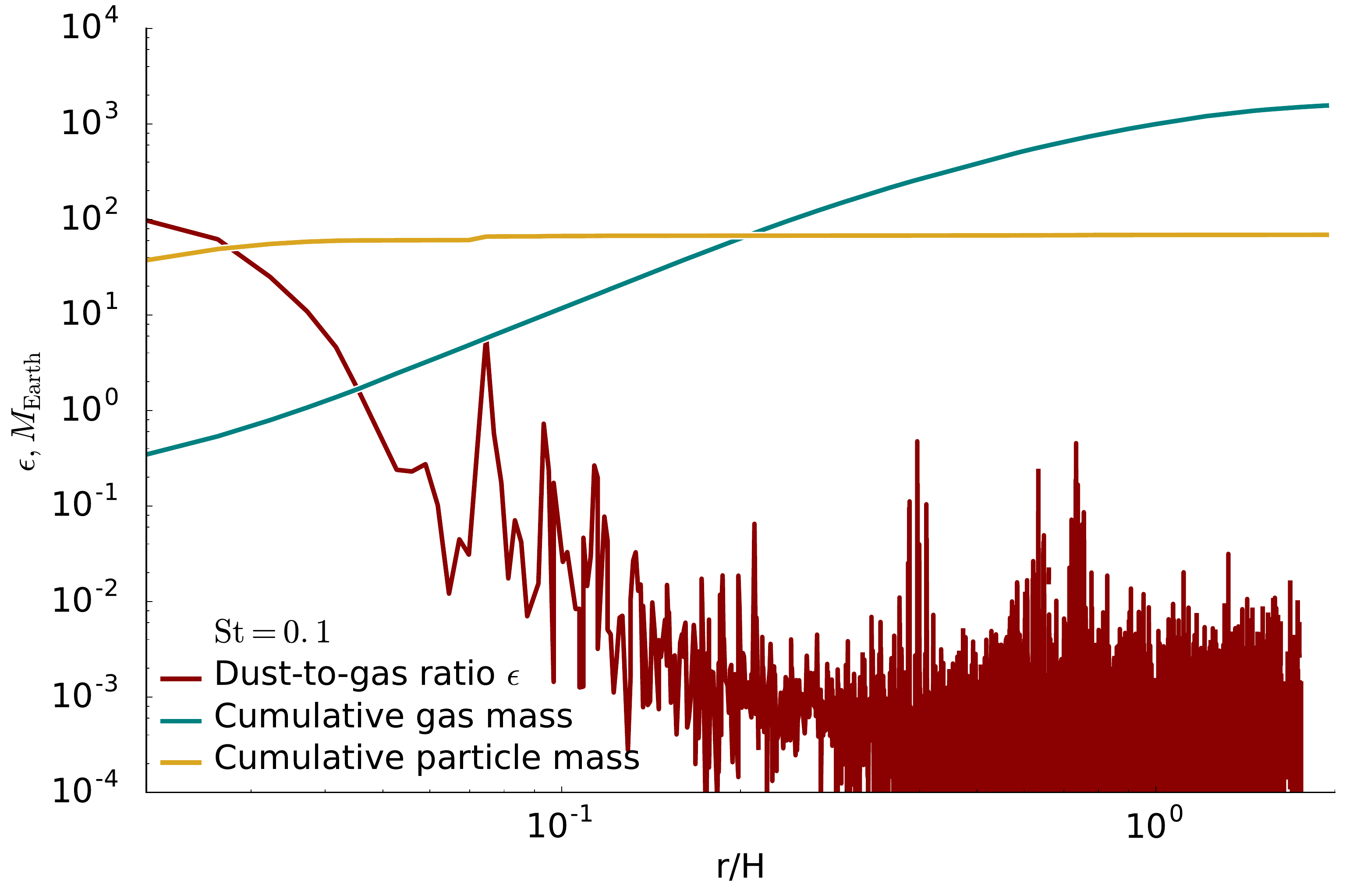}
\includegraphics[width=0.5\textwidth]{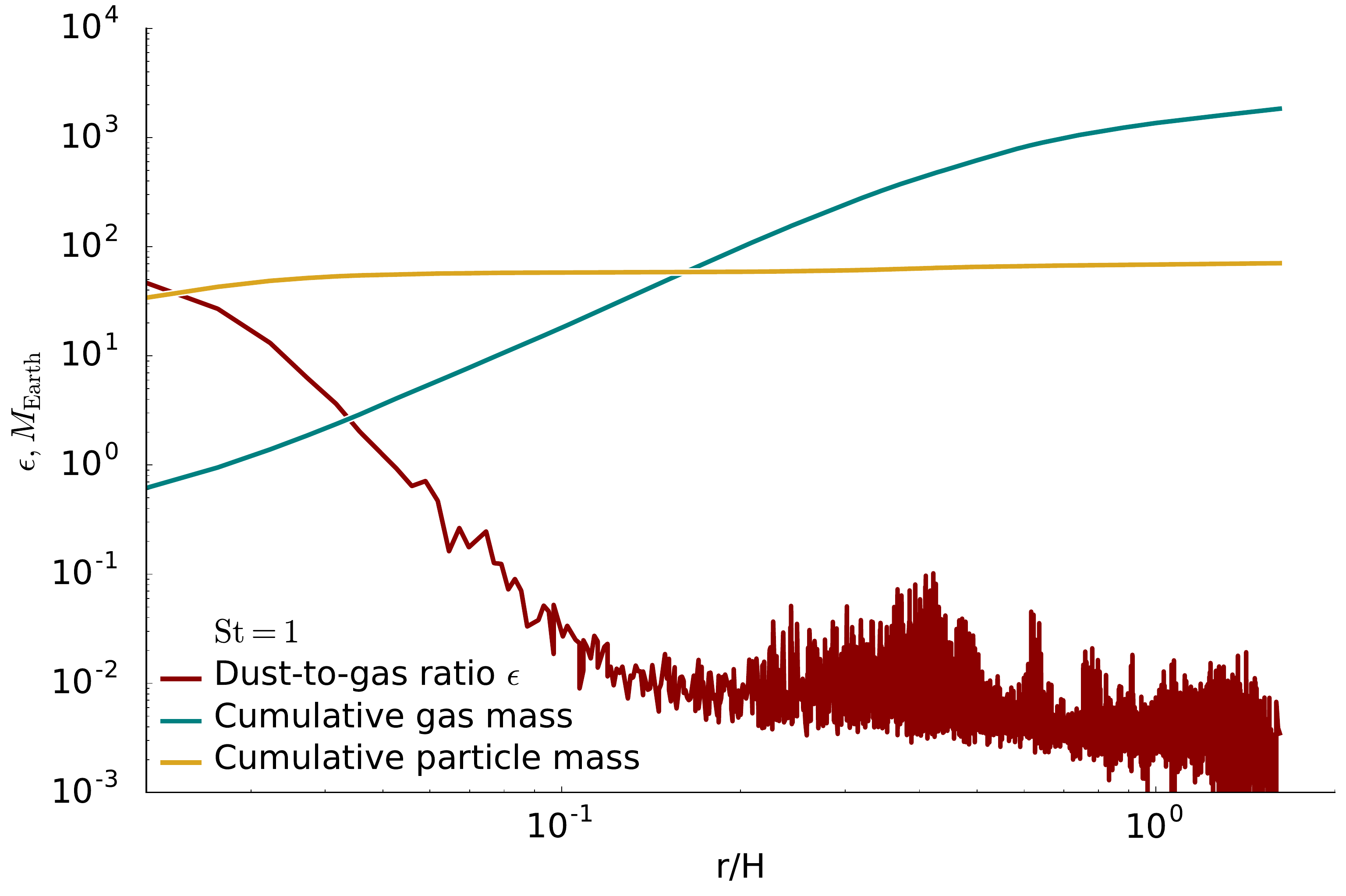}%
\includegraphics[width=0.5\textwidth]{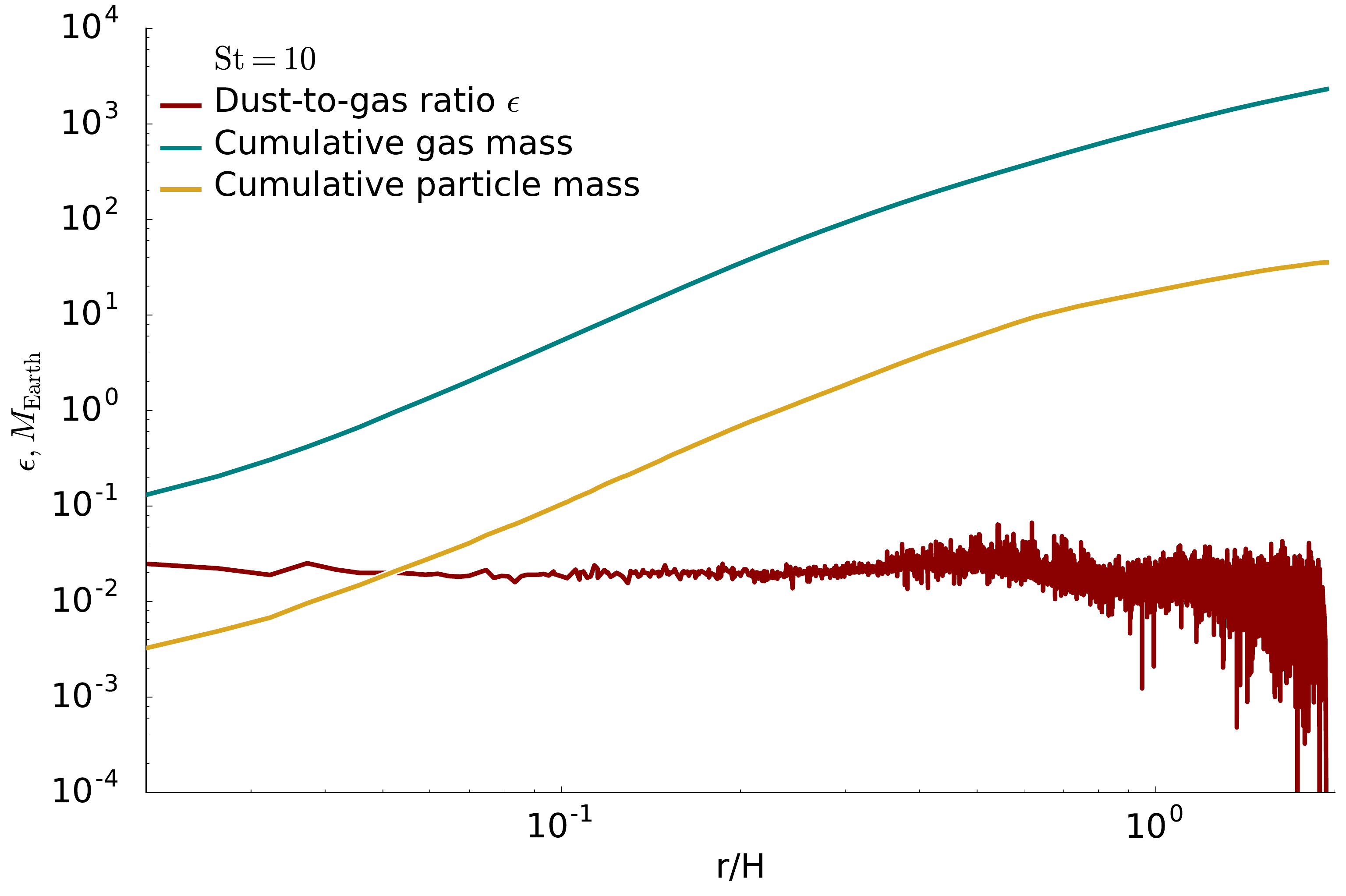}%
\caption{The cumulative gas and solid masses (teal and yellow lines, respectively), summing from the inside and going outwards, of the fragments whose radial mass distributions are shown in Figure \ref{fig:fragmentprofiles}. The red line shows the average dust-to-gas ratio for each radius away from the peak in the gas density. All masses are those of objects expected to be formed at an orbital distance of $100$ au from a $M_{\odot}$ star.}
\label{fig:fragmentmass}
\end{figure*}

Our simulations cover $18 H$ in radial and azimuthal directions and $L_{\rm z} = 2 H$ in the vertical, with a resolution of either 30 or 60 grid points per scale height in all spatial dimensions. This fulfills the resolution requirements to avoid truncation errors which can become unstable \citep{Truelove1997,Nelson2006} as well as the requirements for individual particles, which require that the grid spacing $\Delta x < c_{\textnormal{s}} \tau_{s}$ \citep{Bai2010b}. Additionally, the number of particles must be enough such that at the midplane layer there is 1 particle per cell, which is marginally met in our high grid resolution simulations for a $1024\times 1024$ midplane with $10^{6}$ total particles. Each simulation contains a single particle species of constant size defined by its Stokes number.

Density and temperature are established such that $Q_{\mathrm{0}}$ is on the edge of stability. While \citet{Baehr2017} found that instability requires $Q_{\mathrm{0}} \approx 0.676$, this was likely due to the small radial domain, which excluded the collapse of large unstable wavelengths. Thus for the larger radial-azimuthal domains considered here $Q_{\mathrm{0}} \approx 1$ remains the critical Toomre value, for these simulations meaning that for $c_{\textnormal{s}}=\pi$ and $G=\Omega=1$ the vertically integrated volume density is $\Sigma=1$.

Figure \ref{fig:3d} shows the particle-less fiducial simulation which demonstrates the fragmentation of the gas which is expected in all simulations. Particularly interesting is the presence of a strong coherent anticyclonic vortex at the fragment center as well as a second rotating region separated by a largely stationary gap. If particles concentrate at the center of fragments the existence of a vortex could aid the process.

\section{Results}
\label{sec:results}

Initiated with $Q_{0} = 1$, all simulations quickly proceed to collapse, with particles settling to the midplane at different rates depending on their Stokes number. Smaller particles, i.e. $\mathrm{St} = 0.01$  will take longer as minute gas disturbances will deflect their motion towards the midplane, while larger particles, $\mathrm{St} > 10$, will be able to fall largely unimpeded. Figure \ref{fig:3ddensityevolution} shows the evolution of the maximum gas and particle densities with time for particle sizes (from top to bottom) $\mathrm{St} = [0.1,1,10]$. Gas densities above $\sim 100\rho_{0}$ are considered to be the condition for fragmentation. One can see a clear trend towards higher particle concentrations with smaller particle size.

The rapid concentration of particles within the collapsing axisymmetric and non-axisymmetric structures can be seen in the different panels of Figure \ref{fig:particlemap}. Each panel shows the collapse of the gas and particles for a different particle size, with background color showing the gas surface density with particle surface density overplotted in semi-transparent greyscale. Even though the particles are intended to be uniformly distributed, in practice particles are first even distributed among the processors and then within the grid cells of that processor. This introduces artificial gaps between groups of particles depending on which processor they were originally allocated. These gaps become exaggerated by the large discrepancy between the number of total grid cells and particle number and the shear motions during the linear collapse, apparent in the top left panel of Figure \ref{fig:particlemap}. The particles are not unstable to gravitational collapse when these features are present and thus are not expected to cause any premature fragmentation.

As one might expect, the particles which are most coupled to the gas show the most concentration in the gas overdensities. In the case of our largest particles ($\mathrm{St} = 10$), the inability to couple strongly with the gas results in a steadily decreasing dust-to-gas ratio, unlike in the case of the smaller particles. This is similar to the simulations with 1 km sized rocks from \citet{Boley2010}, which showed the delayed response of the larger particles to the gas drag resulted in wide particle arms which were in different positions than the gas spiral arms. This could lead to scenarios where solids can concentrate outside of the gas overdensities, enriching very different regions of the disk.

Our low-resolution runs showed particle concentration within fragments for particle sizes $\mathrm{St} = [0.01,0.1,1]$, but in the case of $\mathrm{St} = 0.01$, the core is delayed but only because fragmentation of the gas, roughly defined when the maximum gas density is $100\rho_{0}$, is also delayed. This may indicate that there is some dependency of particle fragmentation on the gravitational collapse of the gas, as might be expected from the initial conditions of the gas and dust (gravitationally unstable and stable, respectively).

\begin{figure*}[t]
\centering
\includegraphics[width=0.51\textwidth]{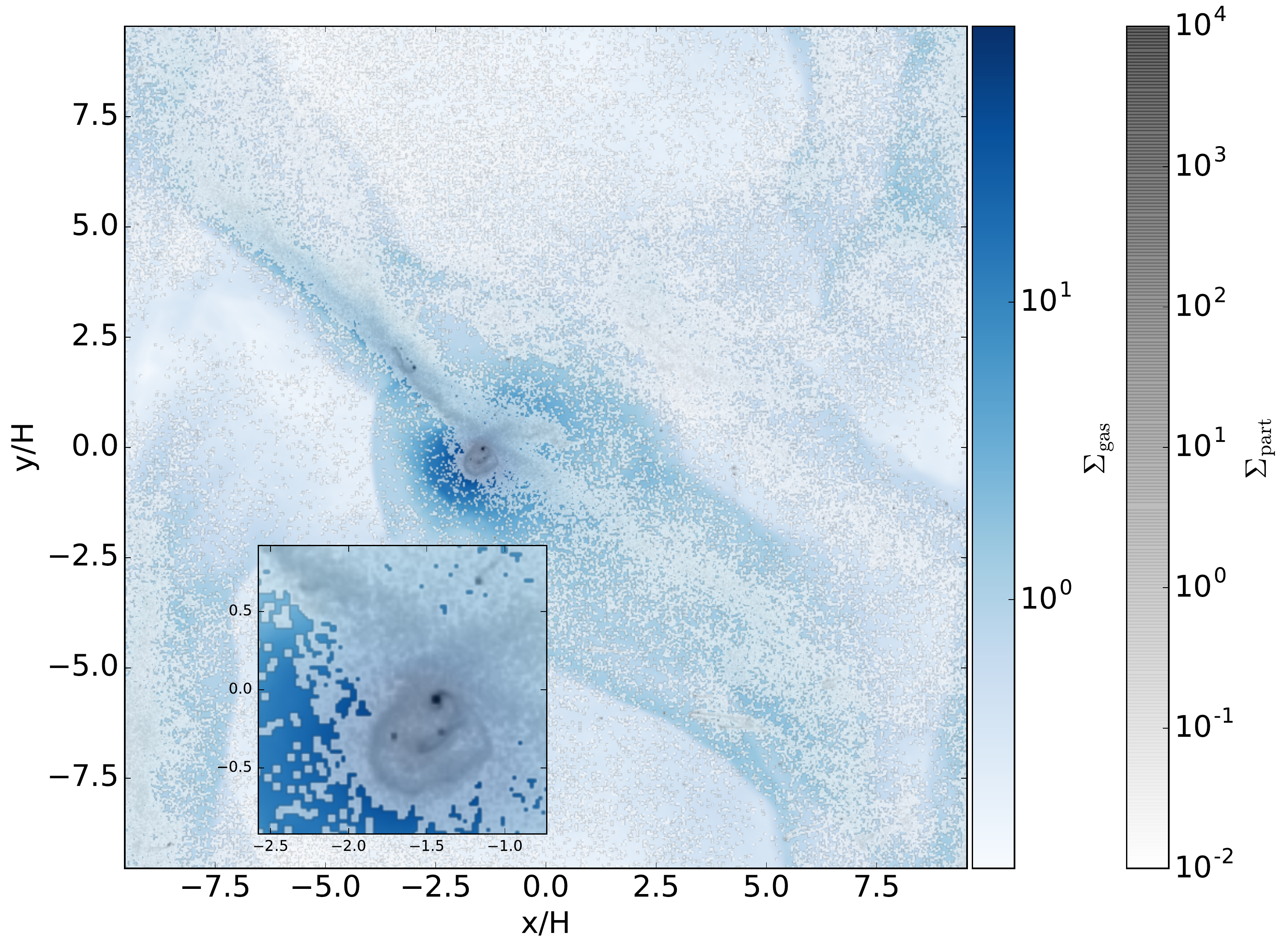}%
\includegraphics[width=0.45\textwidth]{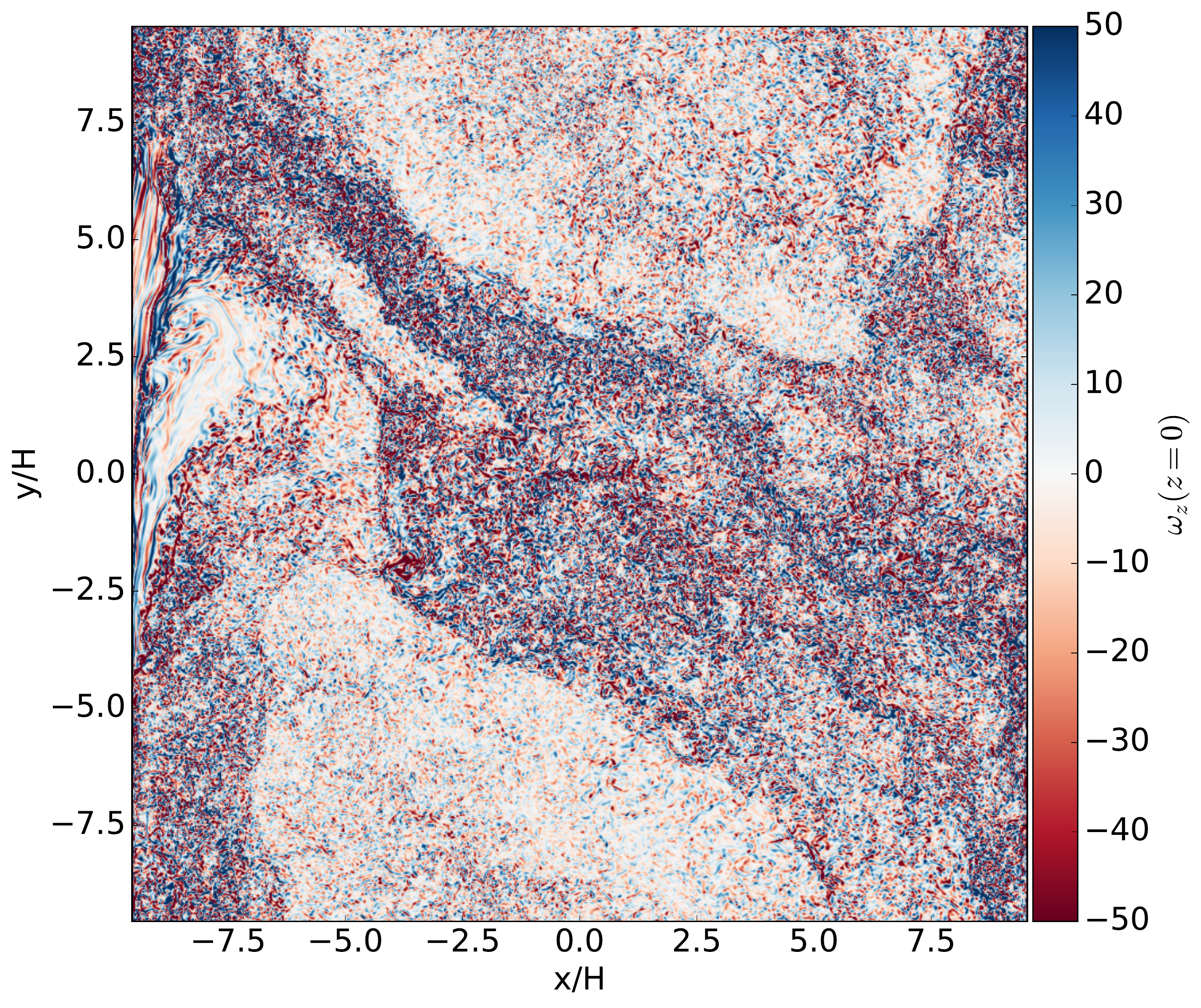}
\caption{A map of the vertically integrated gas and particle densities (left) and a slice at the midplane of the vertical component of the vorticity (right) of a fragmenting 3D simulation. Compared to the Figure \ref{fig:3d}, where a vortex indicates rotation of the fragment and the absence of a vortex with particles included. The inset of the left figure shows a more detailed view of the fragment center and the various particle clusters in its vicinity. This simulation is comparable to Figure \ref{fig:3d}, but with particles and at a higher resolution ($1024^{2} \times 128$).}
\label{fig:3dparticles}
\end{figure*}

In our high-resolution simulations, we again find that small to intermediate particles eventually collect the best within the collapsing fragment and eventually result in at least moderate cores. This is apparent in the radial profiles of the gas fragments which form in our simulations and are shown in Figure \ref{fig:fragmentprofiles}. The gas (teal line) shows a roughly Bonnor-Ebert density profile, with constant density within a radius of $\sim 0.1 H$ and a power law decline thereafter. This is consistent with the isothermal temperature structure assumed for Bonnor-Ebert spheres, here shown with the dark blue line. 

Regardless of the particle size, we still find that the particles will concentrate compared to the background distribution, coinciding with the location of the gas fragment. This is apparent in the top left and bottom right panels of Figure \ref{fig:fragmentprofiles}, which correspond to our largest and smallest particle species, and are least likely to concentrate strongly. While no core is formed, there is still increasing particle density towards the center of the fragment.

In the case of $\mathrm{St} = 0.1$, the core is especially pronounced from the surrounding protoplanetary envelope, able to form a distinct particle clump within the fragment itself. This constitutes the most significant core we form in our simulations and as we detail in the following chapter, is tens of Earth masses in size. The fate of core formation in the simulation \textit{P1024t2pss}, with the highest resolution and smallest particles does not fragment before the timestep becomes prohibitively small, but considering that the lower resolution simulation formed a small but late core, it is reasonable to expect a core is possible in the high resolution case as well.

The simulation \textit{P512t2ps} represents an unusual case where the disk fragments and the particles proceed to collapse into a solid clump, but the gas dissipates shortly thereafter, resulting the peculiarly high metallicity of the clump in that case. Something similar was seen in \citet{Boley2010} where fragments are tidally stripped, leaving a bare core and is speculated as an alternative planet formation pathway, in which tidally stripped cores are free to slowly accrete gas \citep{Nayakshin2010a}. This was however an isolated incident and none of the other simulations were observed to behave in this way.

\subsection{Masses and Solid Concentrations}
\label{subsec:masses}

Importantly, from these simulations we want to determine the size and mass of the solid clumps and how enriched the atmospheres of gas giants or brown dwarfs that form are. First we need to define our initial parameters in physical units consistent with a shearing box located $100$ au from a 1 $M_{\odot}$ star. The orbital frequency at 100 au is $1.98 \times 10^{-10}$ $\mathrm{s}^{-1}$ and the sound speed of a 10 Kelvin monatomic gas (with $\gamma = 5/3$) is $262$ $\mathrm{m\,s}^{-1}$. The Toomre surface density $\Sigma_{\mathrm{T}}$, which is the local surface density required to result in gravitational collapse for a given unstable $Q$ value, can be used with our initial conditions to establish the initial surface density in the simulation 
\begin{equation} \label{eq:toomresurfacedensity}
\Sigma_{\mathrm{T,0}} = \frac{c_{\mathrm{s,0}}\Omega}{\pi G Q_{0}} = 37 \mathrm{\,g\,cm}^{-2}.
\end{equation}
This is roughly an order of magnitude greater surface density than the minimum mass solar nebula density at this radius.

We calculate the masses by determining the unit density at the desired orbital radius and multiplying this value by the total mass in code units of the gas or solids. From $\Sigma_{\mathrm{T}}$, we calculate our unit density from the Gaussian profile of the initial vertical gas mass distribution
\begin{equation} \label{eq:surfacedensity}
\Sigma_{\mathrm{T}} \equiv \int^{H}_{-H} \rho_{\mathrm{0}} e^{-z^{2}/2H^{2}}dz 
\end{equation}
where $\rho_{\mathrm{0}} = 0.29 \hat{\rho}$. Solving for our unit volume density $\hat{\rho}$ gives a value of $5 \times 10^{-12}$ $\mathrm{g\,cm}^{-3}$, with a disk aspect ratio of $H/R = 0.1$. The mass is added up in code units and averaged at all radii around the densest point before being integrated over radial shells and multiplied by $\hat{\rho}$ to calculate the total mass.

The results of these mass calculations, taken at times shortly after the fragmentation, are shown in table \ref{tab:sims}. The profile of the fragments in the table are shown in Figure \ref{fig:fragmentmass}, as the cumulative mass is shown as a function of radius of the fragment $r$, beginning from the outside and going inwards. Since the density of the mass of particles which constitute the solid core fall well short of the density of a solid body of similar mass, around 5 $\mathrm{g\,cm}^{-3}$, these are not true cores at a solid material density, but a cloud of particles at the size of the grid resolution $\sim 0.02 H$. Beyond this point, any further concentration of the particles is limited by the resolution and can be modeled by other means \citep[c.f.][]{Nayakshin2018}.

To get a rough estimate of the maximum initial mass expected for these simulations, one assumes that all mass within a critical wavelength $\lambda_{\rm T} = 2\pi H$
\begin{equation} \label{eq:fragmentinitialmass}
M \approx \lambda_{\rm T}^{2} \Sigma_{\rm T,0} = 17.2 M_{\rm Jup},
\end{equation}
which is the maximum mass our fiducial particleless simulation ever reaches. Since not all material will collapse into the fragment, a lower limit can be approximated by a factor of 4 decrease \citep{Kratter2016}. The majority of the fragments form with initial masses within this $4.3 - 17.2\,M_{\rm Jup}$ range and generally closer to the lower mass end of this range when including particles with back-reaction.

For the purposes of determining the local gravitational instability of the clumps in Figure \ref{fig:fragmentprofiles}, we also define a midplane Toomre volume density $\rho_{T}$, from the 3D Toomre parameter $Q_{\rm 3D}$ of \citet{Mamatsashvili2010}
\begin{equation} \label{eq:toomrevolumedensity}
\rho_{\mathrm{T}} = \frac{\Omega^{2}}{\sqrt{2\pi}\pi G Q_{0}},
\end{equation}
where we have used $\Sigma = \rho H \sqrt{2\pi}$. This sets a threshold midplane mass above which the disk is expected to collapse locally.

For grains in a spherical cloud, sedimentation timescales are inversely proportional to the particle size expressed in terms of the Stokes number. Considering our particles in the Epstein regime, the sedimentation velocity of particles $w$ affected by gravity and drag in a gas of velocity $u$ is \citep{Nayakshin2010a}.
\begin{equation} \label{eq:sedimentationvelocity}
w_{\mathrm{sed}} = w - u = \frac{4\pi}{3}G\rho_{g}R\tau_{s}.
\end{equation}
Assuming the radius of collapse $R$ is equal to the half the Toomre wavelength and the central gas density $\rho_{g}$ to be roughly an order of magnitude greater than the Toomre density, we approximate the sedimentation timescale as
\begin{equation} \label{sedimentationtimescle}
\tau_{\mathrm{sed}} = \frac{\lambda_{\mathrm{T}}}{2(w_{\mathrm{sed}})} = \frac{15\sqrt{2\pi}}{4}\frac{Q_{0}}{\mathrm{St}}\Omega^{-1}.
\end{equation}
This is at odds to the evolution of the three particle species in Figure \ref{fig:3ddensityevolution}, which show that after the gas has fragmented (at around $t = 9\Omega^{-1}$) the particles concentrate to saturated levels considerably faster for the smaller $\mathrm{St}=0.1$ particles, slower for $\mathrm{St}=1$ and not at all for the largest size. This indicates that gravity and aerodynamics are not a sufficient description of the particle collapse and the initial angular momentum and transfer throughout the collapse affects the sedimentation with size. In this case the drift of particle inwards becomes comparable to the circumstellar disks where drift is fastest for particles around $0.1 < \mathrm{St} < 1$\citep{Weidenschilling1977}.

Solid mass ratios of the clumps formed are greater than the initial $Z_{0} = 0.01$ and in the case of $\mathrm{St} \leq 1$, often more comparable to the ratios observed in Jupiter or Saturn. Figure \ref{fig:fragmentmass} shows the total mass in gas (teal) and solids (yellow) as a function of radius for the fragment profiles in Figure \ref{fig:fragmentprofiles}. From this one can see that we consistently form gas fragments of a few Jupiter masses with varying degrees of particle concentration. Particularly in the case of $\mathrm{St} = 0.01$ and $\mathrm{St} = 10$, we can see that although there is several Earth masses worth of material in the entire fragment, it is distributed in a similar fashion to the gas (i.e.: $\epsilon$ remains constant throughout the inner regions of the fragment), and thus no significant core is formed. In the cases of $\mathrm{St} = 0.1$ and $\mathrm{St} = 1$, $\epsilon$ keeps rising as one approaches the fragment center and the formation of a core begins to take shape.

  \subsection{Metallicity}
  \label{subsec:metallicity}

In simulations which do not form a core, the fragment envelopes show radial dust-to-gas ratios consistently larger than their initial $\epsilon_{0} = 0.01$ level, although the scatter is sometimes significant, particularly for values below. Clumps which do form cores have atmospheres depleted in heavier grains, but may have significant peaks in the local dust-to-gas ratio which are the locations of smaller secondary solid bodies which have not immediately accreted to the fragment center. Since later accretion is expected to further alter the C/O ratio by introducing material from chemically distinct regions of the disk, the amount of solid material available at formation is important to understand the observations of gas giant atmospheres \citep{Espinoza2017}. Whether or not a core forms could have a significant impact on the atmospheric metallicity and C/O ratio.

The satellites are perhaps due to the transfer of angular momentum of the particles with the spinning gas fragment. As evident in Figure \ref{fig:3d}, without particles, a fragment forms with significant rotation, mostly cyclonic, but also a weak region of anti-cyclonic rotation. Falling in towards the center of the fragment, the individual particles do not feel the pressure gradient of the gas, but will interact with the rotating gas velocity field. This is borne out not only in the satellites, but also in spiral arms of particles around the fragment center visible in Figure \ref{fig:3dparticles}. These bodies have masses on the order of an Earth mass and remain in an orbit around the primary for the duration of the short simulation runtime.

\begin{figure}[t]
\centering
\includegraphics[width=0.5\textwidth]{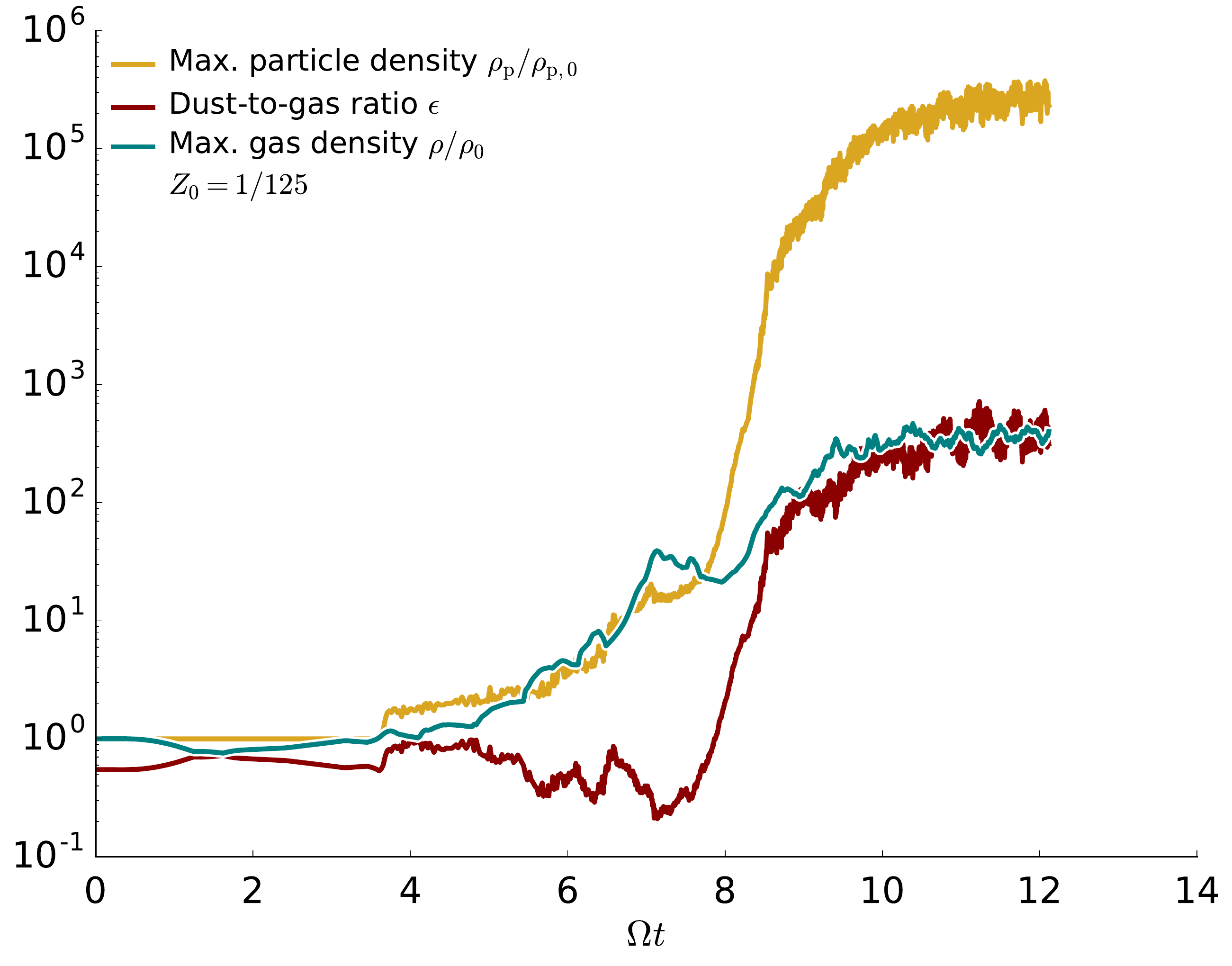}
\includegraphics[width=0.5\textwidth]{density3d1024_ps_axes.pdf}
\includegraphics[width=0.5\textwidth]{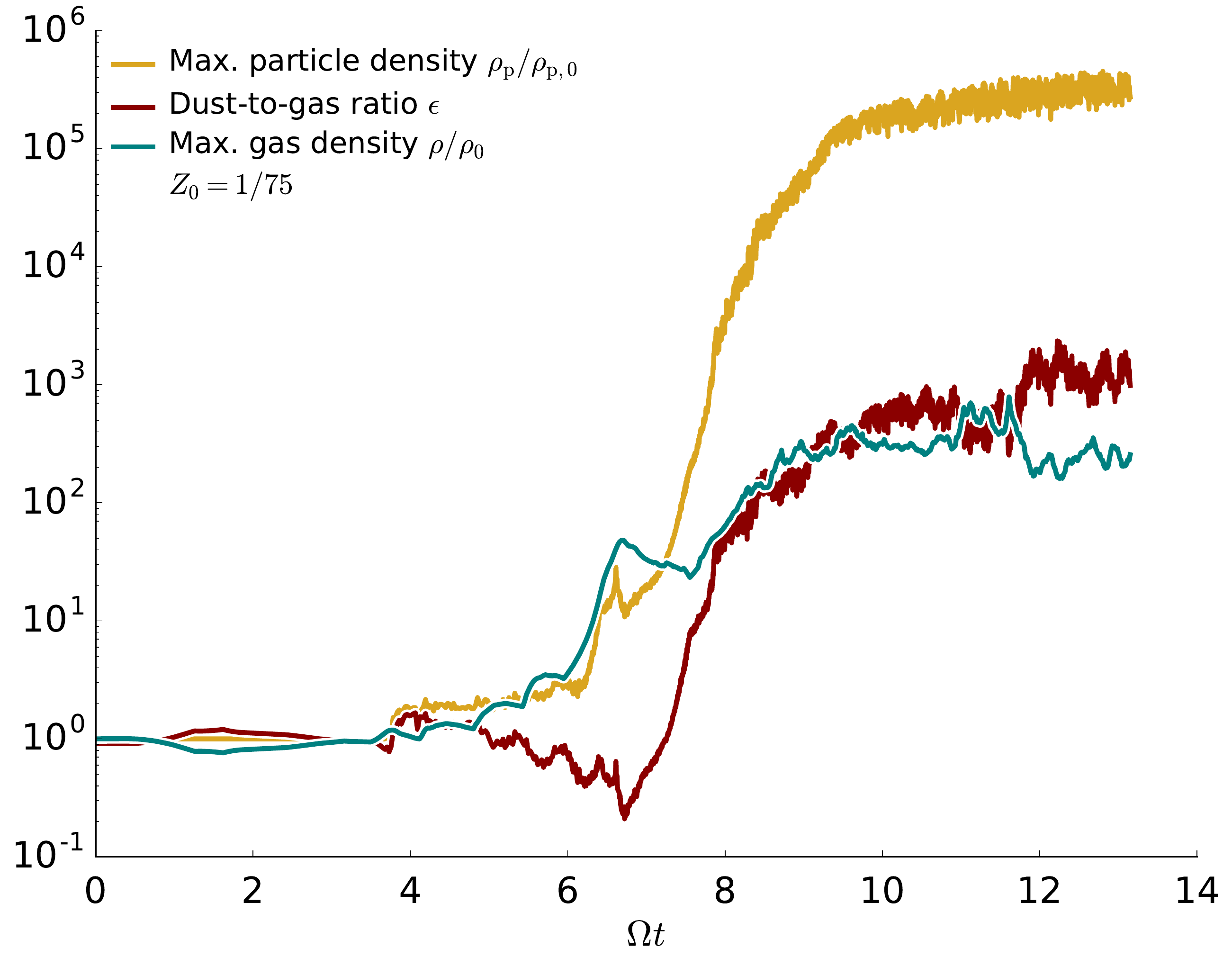}
\caption{Same as Figure \ref{fig:3ddensityevolution}, but with one particle size ($\mathrm{St} = 0.1$) and three different initial dust abundances: $Z_{0}  = 1/125$ (top), $Z_{0} = 1/100$ (middle), $Z_{0} = 1/75$ (bottom). Higher values $Z_{0}$ show faster particle concentration during the non-linear collapse of the gas leading to earlier formation of a massive core, but has little effect on whether or not a core actually forms.}
\label{fig:3ddensityepsilon}
\end{figure}

There is a noted correlation of giant planet occurrence with the stellar metallicity, with observed planets more common around metal-rich stars, a trend which typically supports core accretion planet formation \citep{Fischer2005,Zhu2019}. We added a pair of simulations which deviated from the initial dust abundance to $Z = 0.008$ and $Z = 0.013$ for our simulations which formed the most prominent core, those with particles of size $\mathrm{St} = 0.1$. As seen in Figure \ref{fig:3ddensityepsilon}, the result is a fragment with a core regardless if the initial metallicity is (from top to bottom) $Z_{0} = [1/125, 1/100, 1/75]$. The difference lies in the fragment and core masses, which are significantly higher in both gas and dust components when more solid material is included. Still, the overall metallicity enhancement remains constant for the clumps in all three simulations, with an increase by a factor of around 5. 

\section{Discussion}
\label{sec:discussion}

The concentration of solids in non-fragmenting disks is well-documented \citep{Gibbons2012,Shi2016}, and our simulations show many similar features. For example, intermediate particles sizes $\mathrm{St} = 0.1,1$ concentrate particularly well while smaller and larger particles are either too well coupled to the gas' smallest gas motions or barely influenced by the gas at all, respectively. This is apparent in the non-axisymmetric features of both 2D and 3D simulations, as both cases show most particles concentrating in these structures \citep{Boley2010a,Humphries2018}.

Vortices are known as potential particle traps which could lead to the high dust-to-gas ratios which enable the formation of planetesimals \citep{Raettig2015}. Even in the early, gravitationally unstable disk, vortices have shown the potential to efficiently collect material \citep{Gibbons2015} however vortices may not persist long enough in vertically-stratified simulations \citep{Lin2018a}, affecting the ability of particles to form planetesimals or cores. In Figure \ref{fig:3d}, one can see that our particle-less fiducial simulations fragment and a strong cyclonic vortex spins at its center. However, cyclonic vortices are not good dust traps because since they rotate in the same direction as the disk spins, they will increase the angular momentum of solid material which enter into the vortex, preventing settling.

Theoretical and observational work is coming to the agreement that planetary masses are increasingly rare if formed by gravitational instabilities. Observations have found candidates to be rare occurrences \citep{Vigan2017} and simulations and analytic work have frequently had difficulty forming planetary masses at the right positions and masses and keeping them from migrating away \citep{Baruteau2011a,Kratter2011}. Our simulations here show that initial fragments are already quite large and assuming that they will further accrete material, their masses could very likely increase into the brown dwarf regime. If they do not acquire much further mass the masses here are mostly within the range of many directly observed planets of a few Jupiter masses.

The analysis of \citet{Schlaufman2018} into the relation between giant planets masses and the metallicity of their host stars finds that companions below a mass of $4\,M_{\rm Jup}$ are predominantly found around metal-rich stars while those larger than $10\,M_{\rm Jup}$ do not preferentially form around stars of any metallicity. Thus they find that there are two distinct populations of objects which are separated into planets and brown dwarfs/stars at roughly $10\,M_{\mathrm{Jup}}$. While metallicity does not affect the ability of our simulations to fragment, it does appear to have an effect on the total mass with higher initial metallicities producing fragments with more massive gaseous components. In this respect we find no evidence of a separation between two populations from the initial fragment properties, but that excludes all later dynamics and evolution of the fragment.

In addition to the formation of a core and some satellites, we also notice the formation of a number of smaller solid bodies, particularly in the spiral arms, where particles tend to concentrate when not within the fragment. These objects become much more numerous in the case fragmentation does not occur but instead gravitoturbulence prevails and fragmentation of the dense particle layer only results in many smaller fragmentation events rather than a single blob. Planetesimals such as these may be the seed for a another fragment, contribute to the eventual formation of a terrestrial planet or may be quickly accreted into the fragment and will be the subject of subsequent investigations.

\subsection{Limitations}
\label{subsec:limits}

Many gas-dust interactions have been neglected or simplified in these simulations, including grain heating and cooling, collisional destruction and aggregation, among other effects \citep{Nayakshin2010a}. Collision speeds are expected to be high and may prevent the growth of larger solids, but may be low enough within the spiral arms to allow for aggregation to occur. Additionally, grain growth alone may contribute to the overall Toomre instability of the disk and aid in it's eventual fragmentation \citep{Sengupta2019}. 

\begin{figure}[t]
\centering
\includegraphics[width=0.5\textwidth]{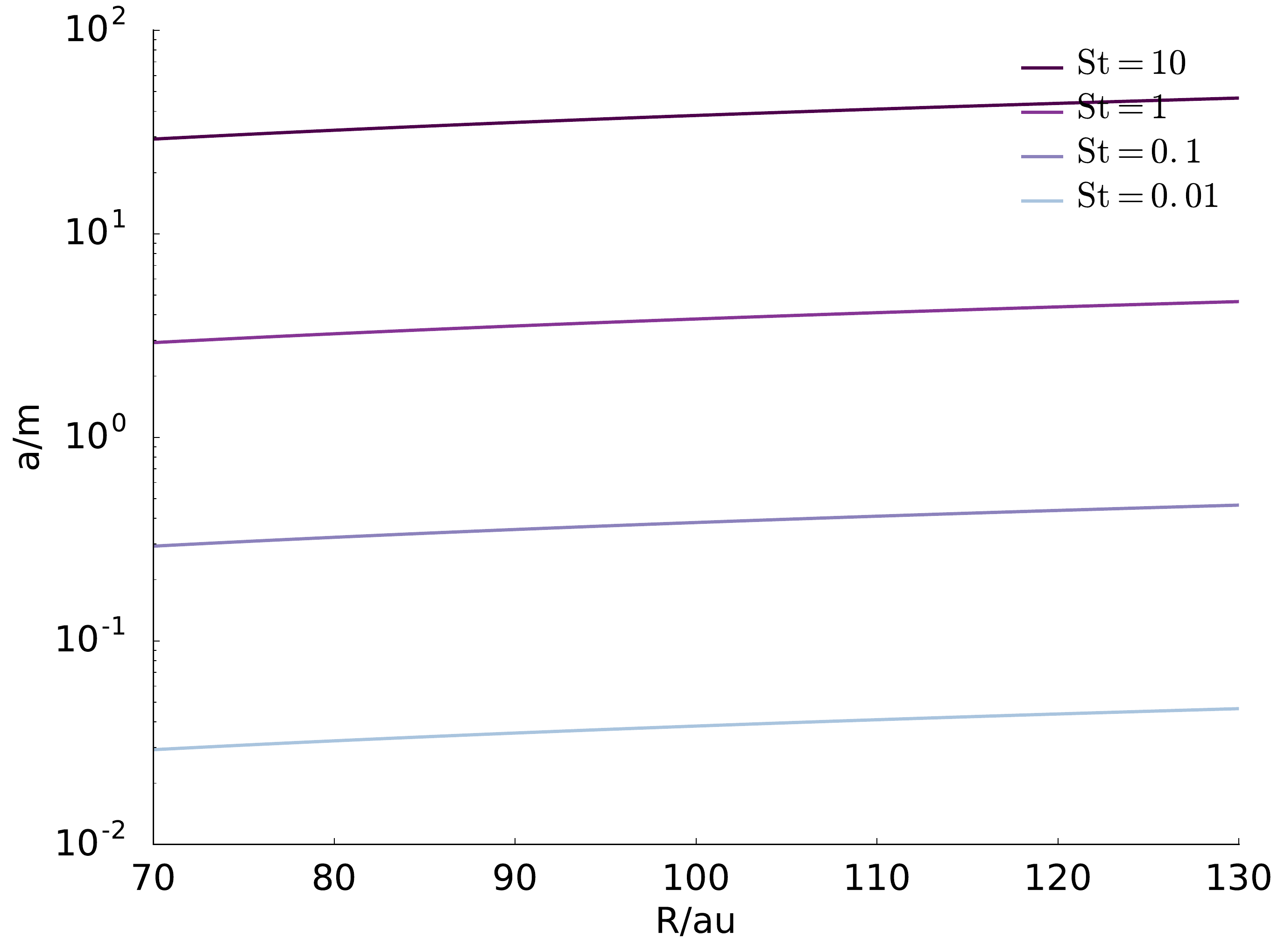}
\caption{The particle sizes to which the Stokes numbers used in this paper correspond, for a range of disk radii around 100 au. Each line assumes a $R^{-1/2}$ dependence of the gas density with radius.}
\label{fig:particlesize}
\end{figure}

Since we use a particle size defined though a Stokes number, the effective size of a particle changes with the properties of the surrounding fluid, which we illustrate in Figure \ref{fig:particlesize}. A particles' size in the Epstein regime scales linearly with the gas density (see Equation \eqref{eq:epsteinregime}) so as a particle moves into a high gas density region, its size will increase which may not necessarily reflect the true nature of the system. This would mean that particles which are originally on the order of centimeters in size would instead have an effective size of meters once inside a gas overdensity one hundred times the initial value. However, as they move through fluid regions of varying density, particles of a particular Stokes number will tend to grow to a mass which maintains their aerodynamic properties \citep{Birnstiel2011}.

Since we do not model the further collapse of a fragment which may significantly increase central temperatures over 1400 K and evaporate the material which constitutes the core \citep{Helled2008}. Additionally, the grains within the clump would begin to dominate the opacity and likely reduce the effectiveness significantly if not completely, rendering the $\beta$-cooling used here lacking. One possible solution which remains computationally inexpensive is to use a modified cooling prescription which drastically reduces cooling efficiency in overdense regions \citep{Baehr2015}. While an already-present core would survive the high temperatures, further growth by dust and pebbles would likely be halted. The long term fate of a core in our simulations will someday benefit from more extensive radiative transfer models which take into account the grain opacities and heating at the fragment centers which could lead to grain evaporation. It should be noted that our cooling prescription includes no metallicity dependence, which is particularly important due to the dominance of dust opacities. For this reason, results might change significantly for a treatment of cooling which takes this into account, especially for the particle-dense central region.

These simulations also assume that the there is a uniform distribution of particle sizes, all with a dust-to-gas ratio of 1:100. This is most certainly not the case, and we expect the smaller particles to be more common since there has been little opportunity for collisional growth and aggregation \citep{Mathis1977}. There is however evidence of some growth in the molecular cloud phase, but with a limit at micron-sized particles \citep{Steinacker2015}, but early gravitoturbulence and particle growth could lead to larger grain sizes \citep{Sengupta2019}. While larger $\mathrm{St}=10$ particles do not collect particularly well within fragments, even $\mathrm{St}=1$ particles may be far less common than smaller species, resulting in less solid accumulation within a fragment.

We do not include any radial pressure gradient, which would result in migration velocities, particularly for $\mathrm{St}=1$ particles, and affect their ability to concentrate at any one location. While these simulations may appropriately describe the local concentration of particles, the absence of a pressure gradient and thus particle drift means that particle concentrations are likely overestimated as particles around $\mathrm{St} = 1$ would migrate quickly away from the gravitationally unstable regions of interest. The spiral arms induced by gravitational collapse are however significant pressure maxima which decrease the particle velocity dispersions and facilitate growth \citep{Booth2016}. Shearing boxes of this size are large enough to have gradients corresponding to global disk properties, but since these simulations are not intended to explore the longer term nature of the fragments that form within we consider the shearing box approximation suitable to model the local collapse and initial conditions of a fragment.

As fragments migrate they may accrete more material, whether it be gaseous or solid, after formation which cannot be captured in such a local model and will affect the overall metallicity \citep{Mercer2017,Ilee2017}. A process similar in principle to pebble accretion \citep{Ormel2010} could result in later metallicity enhancements \citep{Humphries2018}. Fragments are also prone to rapid migration \citep{Baruteau2008} which may take them to new regions of the disk which may be gas or dust rich and accretion will alter their composition which cannot be captured in this study. What is presented in this paper represents a look at the possible initial conditions of gas giant planets formed by GI and their later evolution is a subject of further study.

\section{Conclusions}
\label{sec:conclusion}

We have performed 3D hydrodynamic simulations of gravitationally unstable gaseous disks using high-resolution finite-difference simulations with the {\scshape Pencil} code. Using Lagrangian superparticles within the Eulerian mesh, we modeled the evolution of solids of various sizes in simulations which fragment into dense gas clumps. The overdense arms and fragments are gravitational and hydrodynamic traps for particles which can then concentrate into planetery cores or enrich the metallicity of gas giant planets. We summarize our findings in the following points:

\begin{enumerate}
\item Fragmentation of the gas disk concentrates particles of intermediate sizes initially through aerodynamics, but once concentrated enough they may be comparable gravitationally to the gas fragments themselves. Thus, even though the dust is not initially gravitationally unstable there is still potential for significant solid accumulations to form in fragmenting gravitationally unstable disks, potentially resulting in the growth of solid bodies.
\item Once collected within gas overdensities through aerodynamic drag, solids can concentrate to the point where their contribution to the total self-gravitational potential is non-negligible and thus form considerable core within proto-gas giants. Similar to the results of \citet{Boley2010}, we find that for gas fragments up to several Jupiter masses, with central solid cores typically of a few tens of $M_{\oplus}$.
\item We find that intermediate particle sizes $\mathrm{St=0.1\-- 1}$, fragments form early with already sizable cores before they have core temperatures in excess of $\sim$1400 K which would sublimate solid material. This would mean that sedimentation after formation may not be necessary to explain core in planets formed by gravitational instability.
\item We find fragments in total to have solid fractions are around $1.5 \-- 5\%$, similar to the amounts suggested by core accretion and by observations of Jupiter and Saturn. This suggests that core formation and enhanced metallicity are not unique to core accretion and more information is required of a planet to determine formation mechanism.
\item Our fragments have atmospheres with metallicities above the initial value $Z_{0} = 1/100$ in the case when a core does not form, but below initial metallicities when a core is formed. Whether or not a core forms may have an impact on the atmospheric metallicity of the resulting gas giant planet.
\end{enumerate}

\begin{acknowledgments}
The authors thank the anonymous referee for thorough comments which improved the quality of the paper. HB would like to thank Andreas Schreiber, Chao-Chin Yang, Dominic Batzler and Wlad Lyra for useful discussions. All simulations made use of the Isaac cluster at the Max-Planck Center for Data and Computing in Garching.
\end{acknowledgments}

\bibliography{library}

\end{document}